\documentclass[atlasdraft=false, texlive=2023, UKenglish, cernpreprint, texmf, orcidlogo]{atlasdoc}
\usepackage{atlaspackage}
\usepackage{dcolumn}
\usepackage{atlasbiblatex}

\usepackage{atlasphysics}
\graphicspath{{logos/}{figures/}}

\DeclareUnicodeCharacter{2212}{\textendash}

\usepackage{ANA-HIGG-2018-32-PAPER-defs}
\usepackage{multirow}
\usepackage[force]{feynmp-auto}
\usepackage{multicol}

\AtlasTitle{
Evidence of off-shell Higgs boson production from $ZZ$ leptonic decay channels and constraints on its total width with the ATLAS detector
}

\AtlasAbstract{%
This Letter reports on a search for off-shell production of the Higgs boson using 139 $\textrm{fb}^{-1}$ of $pp$ collision
data at $\sqrt{s}=$ 13~TeV collected by the ATLAS detector at the Large Hadron Collider.
The signature is a pair of $Z$ bosons, with contributions from both the production and subsequent decay of a virtual Higgs boson and the interference of that process with other processes.
The two observable final states are $ZZ\rightarrow 4\ell$ and $ZZ\rightarrow 2\ell2\nu$ with $\ell = e$ or $\mu$.
In the $ZZ\rightarrow 4\ell$ final state, a dense Neural Network is used to enhance analysis sensitivity with respect to matrix element-based discrimination.
The background-only hypothesis is rejected
with an observed (expected) significance of 3.3 (2.2) standard deviations, representing experimental evidence for off-shell Higgs boson production.
Assuming that no new particles enter the production of the virtual Higgs boson, its total width can be deduced from the measurement of its off-shell production cross-section.
The measured total width of the Higgs boson is $4.4^{+3.0}_{-2.2}$~MeV, and the observed (expected) upper limit on the total width is found to be 10.1 (10.8)~MeV at 95\% confidence level.%
}

\AtlasRefCode{HIGG-2018-32}
\PreprintIdNumber{CERN-EP-2023-03}

\AtlasJournalRef{Phys. Lett. B 846 (2023) 138223}
\AtlasDOI{10.1016/j.physletb.2023.138223}

\hypersetup{pdftitle={ATLAS document},pdfauthor={The ATLAS Collaboration}}

\begin{document}

\maketitle
\section{Introduction}
%

% The next lines are included from the .//sections/introduction.tex input file
The discovery of the Higgs boson in 2012 by the ATLAS and CMS collaborations at the Large Hadron Collider (LHC)~\cite{HIGG-2012-27,CMS-HIG-12-028} was a major milestone in particle physics, and since then this particle has been put under the spotlight for further scrutiny to uncover its fundamental nature.
Great progress has been made in measuring the properties and couplings of the Higgs boson~\cite{HIGG-2021-23,CMS-HIG-22-001}, and to date no deviations from the Standard Model (SM) predictions have been found.
The total width of the Higgs boson (\gammaH) is a key prediction of the SM. %
The expected value in the SM (\gammaHSM) for a 125 \GeV~Higgs boson is only 4.1 \MeV~\cite{LHCHiggs:2016ypw},
which is inaccessible via any direct measurement of the width in the resonance region due to limited detector resolution.
To probe this parameter, a method relying on both off-shell and on-shell production of the Higgs boson has been developed, as documented in
~\cite{Kauer:2012hd,Caola:2013yja,Campbell:2013una,Campbell:2013wga}.
In this method, the relationship between the Higgs boson coupling constants in the on-shell and off-shell regimes is assumed to be given by the SM prediction, assuming that no new particles enter into the Higgs boson production process.
On-shell Higgs boson production (only gluon-gluon fusion (ggF) is considered in the equations below, but the principle is the same in other production modes) is inversely proportional to the width:
\begin{equation*}
\sigma^{\text{on-shell}}_{gg\rightarrow H \rightarrow ZZ}\sim~\frac{g^2_{\text{ggF}}g^2_{\text{HZZ}}}{m_H\Gamma_H}.
\end{equation*}
However, off-shell Higgs boson production has no width dependence:
\begin{equation*}
\sigma^{\text{off-shell}}_{gg\rightarrow H \rightarrow ZZ}\sim~\frac{g^2_{\text{ggF}}g^2_{\text{HZZ}}}{m_{ZZ}^2}.
\end{equation*}
Therefore, if the HZZ and effective $ggH$ couplings in the two regimes (where the effective coupling is obtained by treating the quark loop as a single vertex) have a known relationship, \gammaH~can be extracted from the ratio of yields of observed Higgs boson events.
Off-shell production is accessible in the $ZZ$ decay channel because the available phase space for the decay increases rapidly as the off-shell mass approaches the $2m_Z$ threshold, counteracting the expected drop in the production at higher masses
~\cite{Englert:2014aca,Cacciapaglia:2014rla,Azatov:2014jga,Ghezzi:2014qpa,Buschmann:2014sia,Gainer:2014hha,Englert:2014ffa,Goncalves:2017iub,Goncalves:2020vyn,Lee:2018fxj,He:2019kgh,Haisch:2021hvy,Ruhdorfer:2019utl},
where $m_Z$ is the mass of the $Z$ boson.

Multiple searches for off-shell Higgs boson production have been carried out by the ATLAS and CMS collaborations using LHC \RunOne and \RunTwo data~\cite{HIGG-2014-10,CMS-HIG-14-002,CMS-HIG-14-036,CMS-HIG-14-032,HIGG-2017-06,CMS-HIG-18-002,CMS:2022ley}.
In practice the signal of off-shell Higgs boson production is a deficit in $gg\rightarrow ZZ$ or electroweak $q\bar{q}\rightarrow ZZ$ production, due to the negative interference between the off-shell Higgs boson process and the continuum background.
Throughout this Letter, the notation $gg\rightarrow (H^*\rightarrow) ZZ$ is used to refer to the inclusive process that combines the Higgs boson signal $gg\rightarrow H^*\rightarrow ZZ$, the continuum background process $gg\rightarrow ZZ$, and their interference.
Similarly, the notation $q\bar{q}\rightarrow (H^*\rightarrow) ZZ +2j$ refers to the inclusive electroweak process that combines the processes $q\bar{q}\rightarrow H^*\rightarrow ZZ +2j$, $q\bar{q}\rightarrow ZZ +2j$, and their interference.

\begin{figure}
\centering
\subfloat[]{\includegraphics[width=0.24\textwidth]{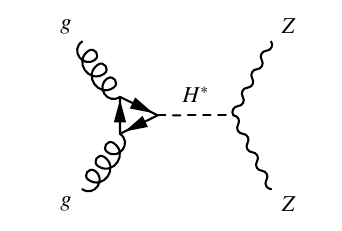}}
\subfloat[]{\includegraphics[width=0.24\textwidth]{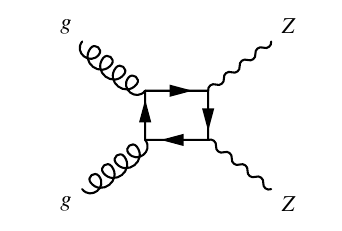}}
\caption{The leading-order Feynman diagrams for the (a) ggZZ signal and (b) background  processes.
In the signal process the quark loop is dominated by top and bottom, while for the continuum background it is mainly light quarks.}
\label{fig:ggZZ_diagrams}
\end{figure}

\begin{figure}[h!]
\centering
\subfloat[]{\includegraphics[width=0.24\textwidth]{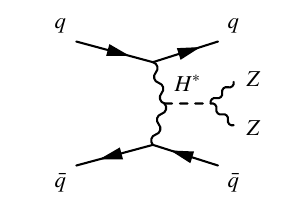}}
\subfloat[]{\includegraphics[width=0.24\textwidth]{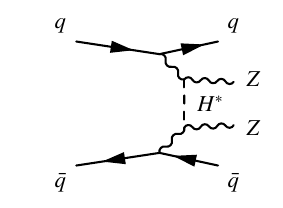}}
\subfloat[]{\includegraphics[width=0.24\textwidth]{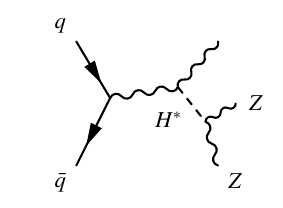}}
\subfloat[]{\includegraphics[width=0.24\textwidth]{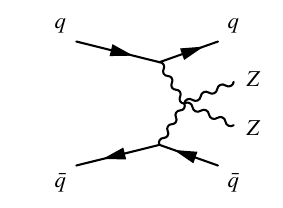}}
\caption{The leading-order Feynman diagrams for (a) the $s$-channel vector-boson fusion signal, (b) the
$t$-channel vector-boson fusion signal, (c) the vector-boson associated production signal, and (d) the vector-boson scattering background.}
\label{fig:EW_diagrams}
\end{figure}

The corresponding leading-order Feynman diagrams for the signal and background processes are shown in Figures \ref{fig:ggZZ_diagrams} and \ref{fig:EW_diagrams}.
Owing to the clean signature and accessible branching fractions, the four-lepton final states (\fourl~and \twoltwov~with $\ell$ $=$ $e$ or $\mu$), originating from the decays of a pair of on-shell $Z$ bosons induced by a virtual Higgs boson, offer the main signal sensitivity.
The latest CMS search~\cite{CMS:2022ley}, using $138~ \textrm{fb}^{-1}$ in the \twoltwov\ channel and $78~ \textrm{fb}^{-1}$ in the \fourl\ channel, led to an observed (expected) detection significance of about 3.6$\sigma$ (2.4) standard deviations ($\sigma$) for off-shell Higgs boson production and a measured \gammaH~of  $3.2^{+2.4}_{-1.7}$ \MeV.
The analysis described in this Letter updates the previous ATLAS result~\cite{HIGG-2017-06} with more data -- the full Run-2 dataset is used in both decay channels -- a more powerful discriminant in the \fourl\ channel, and a data-driven approach to estimating the leading \qqZZ\ background.
Additionally, in this analysis for the first time ggF and EW off-shell production are probed separately as well as together.

This Letter presents a search for off-shell Higgs boson production in four-lepton final states using the full \RunTwo data at a centre-of-mass energy $\sqrt{s}=$ 13 \TeV~collected by the ATLAS detector.
Two decay channels, \fourl~and \twoltwov, are separately analysed and then combined to obtain the final results.
Events with a pair of $Z$ bosons are categorised into several signal regions (SRs) to probe off-shell contributions from the two leading production modes, ggF and electroweak production (EW), and their respective interference with the continuum background $gg\rightarrow ZZ$ and electroweak $q\bar{q}\rightarrow ZZ+2j$ processes.
Electroweak production includes the contributions from vector-boson fusion (VBF) and vector-boson associated production (VH), since these two processes both interfere with the electroweak $q\bar{q}\rightarrow ZZ+2j$ background and hence cannot be separated.
The main irreducible background is $Z$ boson pair production via quark--antiquark annihilation ($q\bar{q}\rightarrow ZZ$); the interfering backgrounds described above also contribute.
In the \fourl~channel, these are the only significant backgrounds, with sub-percent-level contributions from the production of $Z$ bosons with associated jets and \ttbar production.
In the \twoltwov~channel, background processes from diboson production (both $WZ$ and $WW$), \ttbar and single top production, and the production of $Z$ bosons with associated jets constitute roughly half of the total background.
Control regions (CRs) are defined to ensure control of the background modelling.
In both channels, the background from the combination of vector-boson associated production to a top-quark pair ($t\bar{t}$+V, V=W or Z) and triboson production (ZZZ, WZZ, or WWZ) is at the percent level.
Distributions of discriminating variables are fitted simultaneously in all SRs to extract the off-shell contribution by measuring the signal strength \muOFF, the off-shell production cross-section normalised to the SM prediction, with the CRs also included  in the fit to constrain the normalisation of the main background processes.
In the \fourl~channel, an observable is constructed from the output of neural networks (NN) that are trained with kinematic variables and matrix-element discriminants sensitive to the signal process (see Ref.~\cite{HIGG-2017-06}). The \twoltwov~channel uses the transverse mass of the $ZZ$ system,
\begin{equation}
\mtzz \equiv \sqrt{ \left[{\sqrt{\mz^{2} + \left(\pt^{\ell\ell}\right)^2} + \sqrt{\mz^{2} + \left( \met \right)^2}}\: \right]^2 - \left\lvert \vecptll  + \vecmet \right\rvert^2 },
\label{Intro:mTZZ}
\end{equation}
where \mz is the \Zboson boson mass~\cite{ALEPH:2005ab}, $\vecptll$ and $\vecmet$ are the transverse momentum vector of the lepton pair and the missing transverse momentum vector
with magnitudes of $\pt^{\ell\ell}$ and $\met$, respectively.
Finally, the constraint on \gammaH~is derived by using both the measured \muOFF~and the signal strength for on-shell Higgs boson contributions (\muON) in the \fourl~channel obtained from Ref.~\cite{HIGG-2018-29}, relying on the equation (valid under the assumptions discussed above) $\muOFF/\muON=\gammaRatio$. %
Similarly to the previous ATLAS paper~\cite{HIGG-2017-06}, this search also reports the ratio of effective Higgs boson--gluon couplings (\rgg) and the ratio of Higgs boson and vector-boson couplings (\rvv) between the off-shell and on-shell regions, assuming that the Higgs boson total width takes its SM value.

% End of text imported from the .//sections/introduction.tex input file

\FloatBarrier
\section{ATLAS detector}
%

% The next lines are included from the .//sections/detector.tex input file
ATLAS is a multipurpose detector with a forward--backward symmetric cylindrical geometry and a solid-angle\footnote{The ATLAS experiment uses a right-handed coordinate system with its origin at the nominal interaction point (IP) in the centre of the detector and the $z$-axis along the beam pipe. The $x$-axis points from the IP to the centre of the LHC ring, and the $y$-axis points upward. Cylindrical coordinates $(r,\phi)$ are used in the transverse plane, $\phi$ being the azimuthal angle around the $z$-axis. The pseudorapidity is defined in terms of the polar angle $\theta$ as $\eta=-\ln\tan(\theta/2)$.} coverage of nearly 4$\pi$, described in detail in Ref.~\cite{PERF-2007-01}.
The inner tracking detector (ID), covering the region $|\eta| <$ 2.5, consists of a silicon pixel detector, a silicon microstrip detector, and a transition-radiation tracker. The innermost layer of the pixel detector, the insertable B-layer~\cite{Capeans:1291633}, was installed between \RunOne and \RunTwo of the LHC. The inner detector is surrounded by a thin superconducting solenoid providing a $2~\text{T}$ magnetic field, and by a finely segmented lead/liquid-argon (LAr) electromagnetic calorimeter covering the region $|\eta| <$ 3.2. A steel/scintillator-tile hadron calorimeter provides coverage in the central region $|\eta| <$ 1.7. The endcap and forward regions, covering the pseudorapidity range 1.5 $< |\eta| <$ 4.9, are instrumented with LAr electromagnetic and hadron calorimeters, with steel, copper, or tungsten as the absorber material.
A muon spectrometer (MS) system incorporating large superconducting toroidal air-core magnets surrounds the calorimeters. Three layers of precision wire chambers provide muon tracking in the range of $|\eta| <$ 2.7, while dedicated fast chambers are used for triggering in the region $|\eta| <$ 2.4. The trigger system, composed of two stages, was upgraded~\cite{TRIG-PERF} before \RunTwo. The first stage, implemented with custom hardware, uses information from the calorimeters and muon chambers to select events from the 40~MHz bunch crossings at a maximum rate of 100~kHz. The second stage, called the high-level trigger (HLT), reduces the data acquisition rate to about 1~kHz on average. The HLT is software-based and runs reconstruction algorithms similar to those used in offline reconstruction. An extensive software suite~\cite{ATL-SOFT-PUB-2021-001} is used in data simulation, in reconstruction
and analysis of real and simulated data, in detector operations, and in the trigger and data acquisition
systems of the experiment.

% End of text imported from the .//sections/detector.tex input file

\FloatBarrier
\section{Data and Monte Carlo simulation}

% The next lines are included from the .//sections/simulation.tex input file
\label{sec:simulation}

The proton--proton ($pp$) collision data used in this search were collected from 2015 to 2018, corresponding to an integrated luminosity of 139 \ifb. Events in the \fourl~final state were recorded with a combination of single-lepton, dilepton and trilepton triggers, while the \twoltwov~events were collected via multiple single-lepton triggers.
The overall trigger efficiency for the off-shell signal process is more than 98\% in each final state after the application of the SR  selections defined below.

Monte Carlo (MC) simulation is used to predict the normalisation and event kinematics of the signal process and some of the backgrounds.
Event samples for each process were first produced by a corresponding event generator and then passed through detector simulation~\cite{SOFT-2010-01} within the \GEANT~framework~\cite{GEANT4}.
Additional inelastic $pp$ interactions (pile-up) modelled with \PYTHIA{8.186}~\cite{Sjostrand:2007gs} were overlaid on the simulated events to mimic the real collision events, and further corrections were applied to the simulated samples to match the pile-up conditions in the data.
The lepton and jet momentum scale and resolution, and the lepton reconstruction, identification, isolation and trigger efficiencies in the simulation were corrected to match those measured in data.

Separate simulated samples were generated for each of the off-shell signal from ggF production, the $gg\rightarrow ZZ$ background, and the inclusive production $gg\rightarrow (H^*\rightarrow) ZZ$, which also includes the interference between the two.  These loop-induced processes were modelled by \SHERPA~{v2.2.2}~\cite{Bothmann:2019yzt} with \textsc{OpenLoops}~\cite{Buccioni:2019sur,Cascioli:2011va,Denner:2016kdg} at leading-order (LO) accuracy in quantum chromodynamics (QCD), with up to one additional parton in the final state, using the NNPDF3.0 parton distribution function (PDF) set~\cite{Ball:2014uwa}.
Signal and background were simulated separately from the inclusive process for NN training and template fitting.
The merging with the parton shower was performed using
the MEPS@NLO prescription~\cite{Cascioli:2013gfa} and the \SHERPA{} built-in algorithm was used for parton showering and hadronisation.
The samples are corrected to next-to-leading order (NLO) in QCD using corrections calculated separately as a function of the invariant mass of the $ZZ$ system (\mzz)~\cite{Caola2015,Caola2016} for the signal, background, and inclusive processes.
These corrections are similar for each process and range from 1.5 to 2.
The total normalization of all three processes was then corrected to next-to-next-to-next-to-leading order (N3LO) in QCD using a constant correction of 1.32, derived for the off-shell signal~\cite{Passarino:2013bha, YR4}.
The use of the same correction for all processes is justified as the N3LO corrections are expected to be very similar for signal, background, and interference~\cite{PhysRevD.88.034032,CAOLA2016275}.

EW production of $ZZ$ and two jets, also denoted by $q\bar{q}\rightarrow (H^*\rightarrow)~ZZ +2j$, contains inclusively the off-shell signal from VBF production, VH production, the non-Higgs boson EW $ZZjj$ process, and their interferences. Those processes were modelled by \MGNLO~\cite{Alwall:2014hca} at LO QCD accuracy using the NNPDF3.0 NLO PDF set~\cite{Ball:2012cx}. \PYTHIA{8.244} \cite{Sjostrand:2007gs} was used for parton showering and hadronisation %
with the A14 set of tuned parameters (tune) for the underlying event~\cite{ATL-PHYS-PUB-2014-021}.
The $t$-channel exchange of the Higgs boson is treated as a contribution to the VBF signal process.

The ggF- and VBF-induced contributions can be straightforwardly parameterised as a function of \muOFF, where the off-shell signal and the interference depend on \muOFF\ and $\sqrt{\muOFF}$, respectively. More details of this parameterisation are given in Section \ref{sec:results}.

The $q\bar{q}\rightarrow ZZ$ background was simulated by \SHERPA~{v2.2.2} with \textsc{OpenLoops} using the NNPDF3.0 NNLO PDF set. The matrix elements were calculated to NLO accuracy in QCD for 0- and 1-jet final states, and to LO accuracy for 2- and 3-jet final states. The merging with the \SHERPA parton shower was performed using the MEPS@NLO prescription~\cite{Hoeche:2012yf}.
NLO EW corrections calculated on top of the LO QCD prediction were applied as a function of \mzz~for the \fourl~final state~\cite{Biedermann:2016yvs,Biedermann:2016lvg}, while the \mzz-based corrections for the \twoltwov~channel were averaged from the additive (NLO EW + NLO QCD) and multiplicative (NLO EW $\times$ NLO QCD) approaches following Ref.~\cite{Kallweit:2017khh}.
A cross-check was performed in the \fourl~channel, and the results of the two approaches were found to agree within their uncertainties, while the uncertainties of both approaches (described in detail below) were also found to be similar.

The $WZ$ diboson events from both QCD and EW production,
with the subsequent leptonic decays of both the $W$ and $Z$ bosons, were simulated using \SHERPA with a similar set-up to that of the $q\bar{q}\rightarrow ZZ$ background.
The $WZ$~events with the $Z$ boson decaying leptonically and the $W$ boson decaying hadronically were modelled with \SHERPA~{v2.2.1}.
For the \twoltwov~final state, the contribution from $WW$ production was removed in the \SHERPA{} simulation of the $q\bar{q}\rightarrow ZZ$~and $gg\rightarrow ZZ$~processes by requiring the charged leptons and the neutrinos to have different lepton flavours (the prediction was then scaled up by 1.5 to compensate).
The $q\bar{q}\rightarrow WW$ and $gg\rightarrow WW$ processes were then modelled with \POWHEGBOX~{v2}~\cite{Nason:2013ydw} and \SHERPA~{v2.2.2}, respectively.
The interference between $WW$ and $ZZ$ production in the \twoltwov\ final state is expected to be negligible in the phase space of the analysis~\cite{Kallweit:2017khh} and was therefore not considered.

The $Z+$jets background was simulated using the \SHERPA~{v2.2.1} event generator, where the matrix elements were calculated for up to two partons at NLO and four partons at LO. The $Z+$jets events were normalised using the NNLO cross-sections~\cite{GAVIN20112388}.
The $t\bar{t}$ background, as well as single-top (including s-channel, t-channel, and the dominant $Wt$ component) production, were modelled using \POWHEGBOX{v2} interfaced to \PYTHIA{8.230} with the A14 tune.
The total cross-sections for $t\bar{t}$ production and single-top production were normalised to the predictions at NNLO and NLO accuracy in QCD~\cite{Czakon:2011xx,Kant:2014oha,Aliev:2010zk}, respectively.

The triboson backgrounds $ZZZ$, $WZZ$, and $WWZ$ with fully leptonic decays were modelled  with \SHERPA~{v2.2.2} at NLO QCD accuracy.
The $ZZZ\rightarrow 4\ell +2j$ process is included in the $q\bar{q}\rightarrow (H^*\rightarrow)~ZZ +2j$ sample described above.
The simulation of $t\bar{t}+V$ production ($V =$ $W$ or $Z$) with at least one of the top quark decaying leptonically and the vector boson decaying inclusively was performed with \MGNLO~interfaced to \PYTHIA{8.210} for parton showering and hadronisation
with the A14 tune. The total cross-sections for the $t\bar{t}+V$ backgrounds were normalised to the NLO QCD and EW predictions from Ref.~\cite{Frixione:2015zaa}.

% End of text imported from the .//sections/simulation.tex input file

\FloatBarrier
\section{Reconstruction of physics objects}
%

% The next lines are included from the .//sections/object.tex input file
To describe the event signature and obtain a good signal-to-background ratio, this search relies on the successful reconstruction of collision vertices, electrons, muons, jets, $\vec{E}_{\mathrm{T}}^{\mathrm{miss}}$, as well as identification of jets containing $b$-hadrons ($b$-jets). The reconstruction is identical to that in Ref.~\cite{HIGG-2018-09}, and briefly summarised as follows.

Events are first required to have a collision vertex associated with at least two tracks each with transverse momentum \pt~$>0.5$~\GeV. The vertex with the highest sum of $p_{\mathrm{T}}^{2}$ of the associated tracks is referred to as the primary vertex.

Muons are primarily identified by tracks or segments (tracks using the hits of a single MS station) reconstructed in the MS and matched to tracks reconstructed in the ID, with exceptions in areas where the MS lacks coverage.
In the region $2.5<|\eta|<2.7$, muons can also be identified by tracks from the muon spectrometer alone (standalone muons).
In the gap region ($|\eta|<0.1$) of the MS, muons can be identified by a track from the ID associated with a compatible calorimeter energy deposit (calorimeter-tagged muons).
Candidate muons are required to have \pt $> 5$ \GeV~and $|\eta| < 2.7$, with the exception of calorimeter-tagged muons for which the \pt threshold is raised to 15~\GeV.
Muons must satisfy the `loose' identification criterion~\cite{PERF-2015-10} in the \fourl~channel with at most one standalone or calorimeter-tagged muon allowed per Higgs boson candidate.
In the \twoltwov~channel muons are selected with $|\eta| < 2.5$ and must satisfy the `medium' identification criterion.
Electrons are reconstructed from energy deposits in the electromagnetic calorimeter matched to a track in the ID. Candidate electrons must have $\pT > 7$~\GeV{} and $|\eta| < 2.47$, and satisfy the `loose' and `medium' identification criteria~\cite{PERF-2017-01} in the \fourl~and \twoltwov~channels, respectively.
All electrons and muons used in both channels must be isolated and satisfy the `FixedCutPFlowLoose' isolation criteria~\cite{PERF-2015-10,PERF-2017-01}.
Furthermore, electrons (muons) are required to have associated tracks satisfying $|d_0/\sigma_{d_0}|<5~(3)$ and $|z_0\times\sin\theta|<0.5$~mm, where $d_0$ is the transverse impact parameter relative to the beam line, $\sigma_{d_0}$ is its uncertainty, and $z_0$ is the z coordinate of the r-$\phi$ impact point, defined relative to the primary vertex. The event is rejected if the minimum angular separation between two leptons is \drll$< 0.1$, where \drll~$=\sqrt{(\Delta\phi_{\ell\ell})^2 + (\Delta\eta_{\ell\ell})^2}$.

Jets are reconstructed from particle-flow objects~\cite{ATLAS:2017ghe} using the anti-$k_t$ algorithm~\cite{antikt_algorithm,Fastjet} with radius parameter $R = 0.4$. The jet-energy scale is calibrated using simulation and further corrected with in situ methods~\cite{Aaboud:2017jcu}.
Candidate jets are required to have \pt $>$ 30 \GeV~and $|\eta| <$ 4.5.
A jet-vertex tagger~\cite{PERF-2014-03} is applied to jets with $\pT<60$~\GeV\ and $|\eta|<2.4$ to suppress jets that originate from pile-up.
To mitigate the impact of pile-up jets in the forward region, another tagger, based on jet shapes and topological jet correlations~\cite{ATLAS:2017ywy}, is used to suppress jets originating from the pile-up with $\pT<50$~\GeV\ and $2.5<|\eta|<4.5$.
In addition, $b$-jets are identified using a multivariate $b$-tagging algorithm~\cite{Aad:2019aic} and events containing them are rejected.
The chosen $b$-tagging algorithm has an efficiency of 85\%{} for $b$-jets and a rejection factor of 33 against light-flavour jets, measured in $t\bar{t}$ events \cite{ATLAS:2022qxm}.

The presence of neutrinos is identified using the missing transverse momentum vector $\vec{E}_{\mathrm{T}}^{\mathrm{miss}}$, which is computed as the opposite of the vector sum of transverse momenta of all the leptons and jets, as well as the tracks originating from the primary vertex but not associated with any of the leptons or jets~\cite{ATL-PHYS-PUB-2015-027}.
Missing transverse momentum may also arise from the mismeasurement of the momentum of particles or jets.
To avoid accepting events due to the presence of this kind of fake \met, the statistical significance of the \met, S(\met), is used.
S(\met) is calculated from the resolution information of the physics objects used in the \met~reconstruction~\cite{ATLAS-CONF-2018-038}.
%

% End of text imported from the .//sections/object.tex input file

\FloatBarrier
\section{\zzfourl\ analysis}

% The next lines are included from the .//sections/analysis4l.tex input file
\label{sec:analysis4l}

The selection of candidate events used in the signal and control regions of the \fourl\ channel closely follows that described in Ref~\cite{HIGG-2018-09}. The four-lepton invariant mass is required to be above the on-shell $ZZ$ production threshold, \mfourl~ $>$ 180 \GeV.
Candidate \fourl\ quadruplets are formed by selecting two opposite-sign, same-flavour dilepton pairs in each event.
In the 4e and 4$\mu$ channels, in which there are two possible pairings, the one that includes the lepton pair with mass closest to the Z boson mass is chosen.
The \pT\ thresholds for the three leading leptons are 20, 15 and 10 \GeV, respectively. In each quadruplet, the lepton pair with mass closest to the $Z$ boson mass, $m_{12}$, is referred to as the leading pair and required to have $50 <m_{12} < 106$ \GeV.
The sub-leading pair, $m_{34}$, must satisfy $50 <m_{34} < 115$~\GeV~when $\mfourl>190$ \GeV.
Due to the increased probability of one off-shell Z boson at lower values of \mfourl, the lower threshold for $m_{34}$ decreases linearly to 45 \GeV\ for $180< \mfourl < 190$ \GeV.

Three SRs are designed to provide sensitivity to both the EW and ggF production modes.
The SRs are defined such that \mfourl\ is well above the Higgs boson mass, including only events with $\mfourl>220$ \GeV.
Events in the range $180<\mfourl<220$~\GeV\ are expected to have the lowest signal-to-background ratio in the \mfourl\ range of the analysis and so are reserved for the control regions defined below.
Events containing two or more jets with \pT greater than 30 \GeV, where the two leading jets are well separated in $\eta$, $|\detajj| > 4$, are classified into the EW SR.
Events falling outside the EW SR but featuring exactly one jet in the forward direction ($|\eta_j|>2.2$) are assigned to a mixed SR.
All the remaining events are then assigned to the ggF SR.

The main background in the \fourl\ channel is the $q\bar{q}\rightarrow ZZ$ process. The overall normalisation of this background is constrained by data in three different CRs defined with $180<\mfourl <220$ \GeV~and with zero, one, or $\geq 2$ jets.
The signal contamination in these CRs is below 2\%.
The kinematic distributions are modelled with simulation, described in Section \ref{sec:simulation}.
Events in the zero- and one-jet CRs are binned in four and two intervals of equal width in \mfourl, respectively, to provide further information about event kinematics.
The interfering background processes $gg\rightarrow ZZ$ and EW $q\bar{q}\rightarrow ZZ$, as well as the small backgrounds from triboson production and $t\bar{t}Z$, are estimated from simulation.
The contribution of the reducible backgrounds where hadrons or their decay products are mis-reconstructed as prompt leptons, such as $Z$+jets, $WZ$ and $t\bar{t}$ processes, are estimated by using data-driven methods described in Ref.~\cite{HIGG-2018-09} and found to be negligible.

To maximize the signal sensitivity, a multi-class dense NN is employed in the SRs to enhance events with a Higgs boson candidate.
The NN, implemented using Keras~\cite{chollet2015keras}~with TensorFlow~\cite{tensorflow2015-whitepaper}~as the backend, is designed to differentiate among the three event classes: the off-shell Higgs boson signal (S), the interfering background (B), and the non-interfering (NI) background.
The interfering backgrounds to the ggF and EW signals are the $gg\rightarrow ZZ$ and EW $q\bar{q}\rightarrow ZZ+2j$ processes, respectively. The non-interfering background is the $q\bar{q}\rightarrow ZZ$ process in both production modes.

The outputs of the NN use a normalized exponential function so that they can be interpreted as probabilities of an event belonging to a particular class ($P_\textrm{S}$, $P_{\textrm{B}}$ and $P_{\textrm{NI}}$) and their ratio is used to define the final observable:
\begin{equation*}
O_\textrm{NN} = \log_{10}\left(\frac{P_\textrm{S}}{P_{\textrm{B}}+P_{\textrm{NI}}}\right).
\end{equation*}
As the analysis attempts to constrain both the ggF- and EW-induced off-shell signals independently, two separate NNs are trained, one in the ggF SR and the other in the EW SR. The observable from the first NN (\ONNggF) is then used as the discriminating variable in both the ggF and mixed SRs, while that of the second NN (\ONNVBF) is used in the EW SR.

The first NN is trained to discriminate among the ggF-induced signal, the $gg\rightarrow ZZ$ background, and the $q\bar{q}\rightarrow ZZ$ process.
The features used by this NN include the kinematic information of the four leptons from MC simulation and also the square of the modulus of the values of the LO matrix element (ME) for the four leptons.
The LO MEs are calculated for the gluon-induced signal and background processes and the \qqZZ\ process from the final-state variables in the Higgs boson rest frame using the \textsc{MCFM} program~\cite{Campbell:2013una,HIGG-2017-06}.
The kinematic variables are the leading $Z$ boson production angle and four decay angles defined in Ref~\cite{HIGG-2013-01}, the three invariant masses \mfourl, $m_{12}$ and $m_{34}$.
These are used as inputs to the ME calculation, and, along with the transverse momentum of the four-lepton system, as inputs to the NN as well.

The second NN is used to separate the EW-induced off-shell signal process from the non-Higgs boson EW $q\bar{q}\rightarrow ZZjj$ background and the QCD-induced $q\bar{q}\rightarrow ZZjj$ process.
In addition to the variables used in the first NN, with matrix elements calculated specifically for the final state with two jets, several supplementary variables are included to exploit the kinematics of the dijet system: the invariant mass and azimuthal separation of the two leading jets, and the two Zeppenfeld angular variables, calculated for each Z boson as $\eta_{\text{Zep}}=\eta_{Z_1}-(\eta_{j_1}+\eta_{j_2})/2$~\cite{PhysRevD.54.6680}.

The two networks have 7 and 9 hidden layers respectively, with [90, 80, 80, 75, 75, 40, 40] and [60, 65, 70, 85, 90, 80, 75, 50, 30] neurons.
The sparse categorical cross-entropy loss was used to optimize the network structure as well as the learning rate of the Adam optimizer.
Input features were chosen to fully describe the event kinematics.
The networks were trained on samples of the signal and the interfering and non-interfering backgrounds.

Figure~\ref{analysis4l:PostFit_SRs_main} shows the distributions of the observed and expected NN-based observables in all three signal regions.
The data is shown after the simultaneous fit described in Section \ref{sec:results}, except that the fit is carried out only in the \fourl\ channel and the value of $\muOFF$ is set equal to one.
All systematic uncertainties, which are described in Section~\ref{sec:systematics}, are included.

\begin{figure}[h!]
\centering
\subfloat[]{
\includegraphics[width=0.48\textwidth]{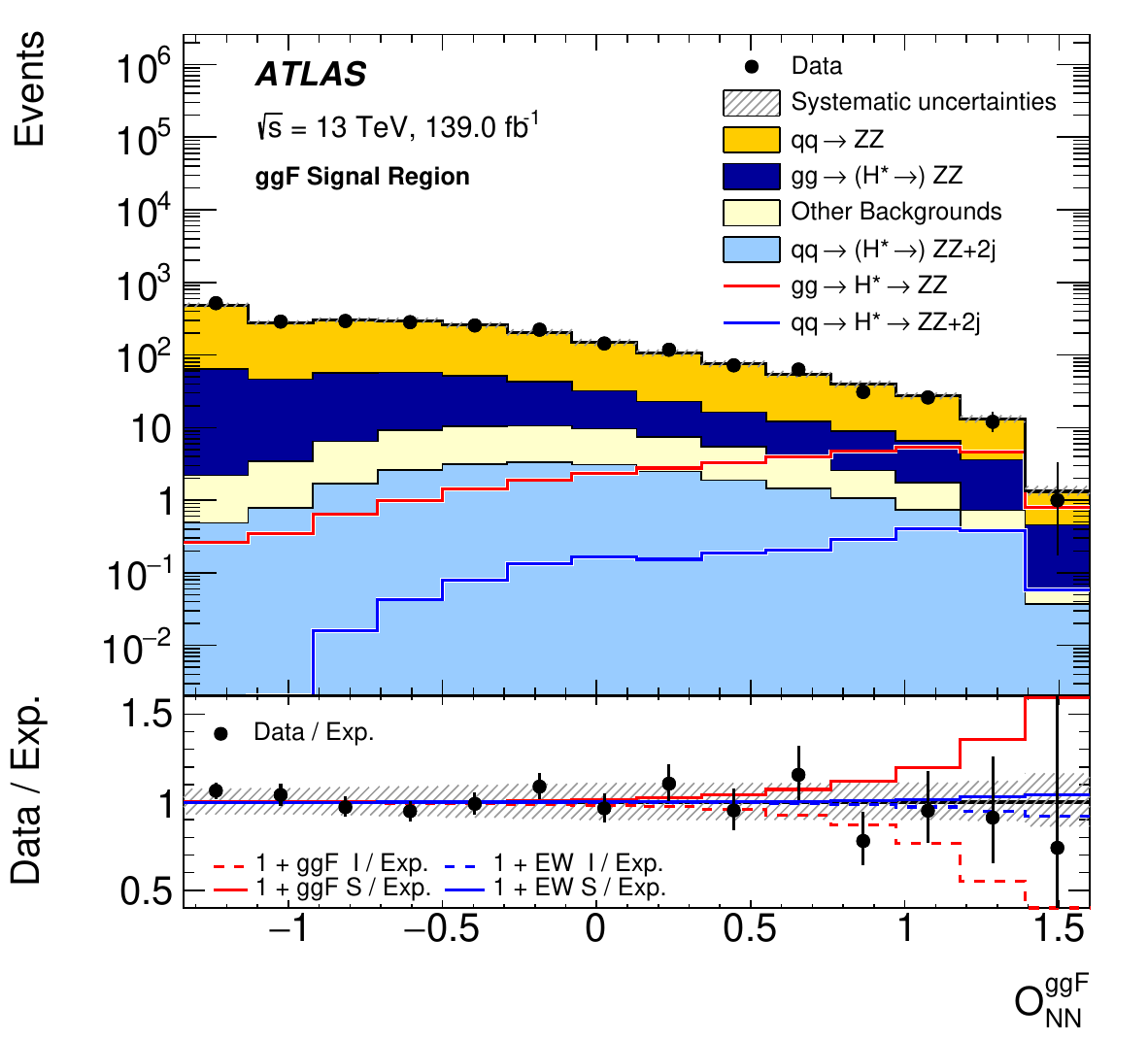}
}
\subfloat[]{
\includegraphics[width=0.48\textwidth]{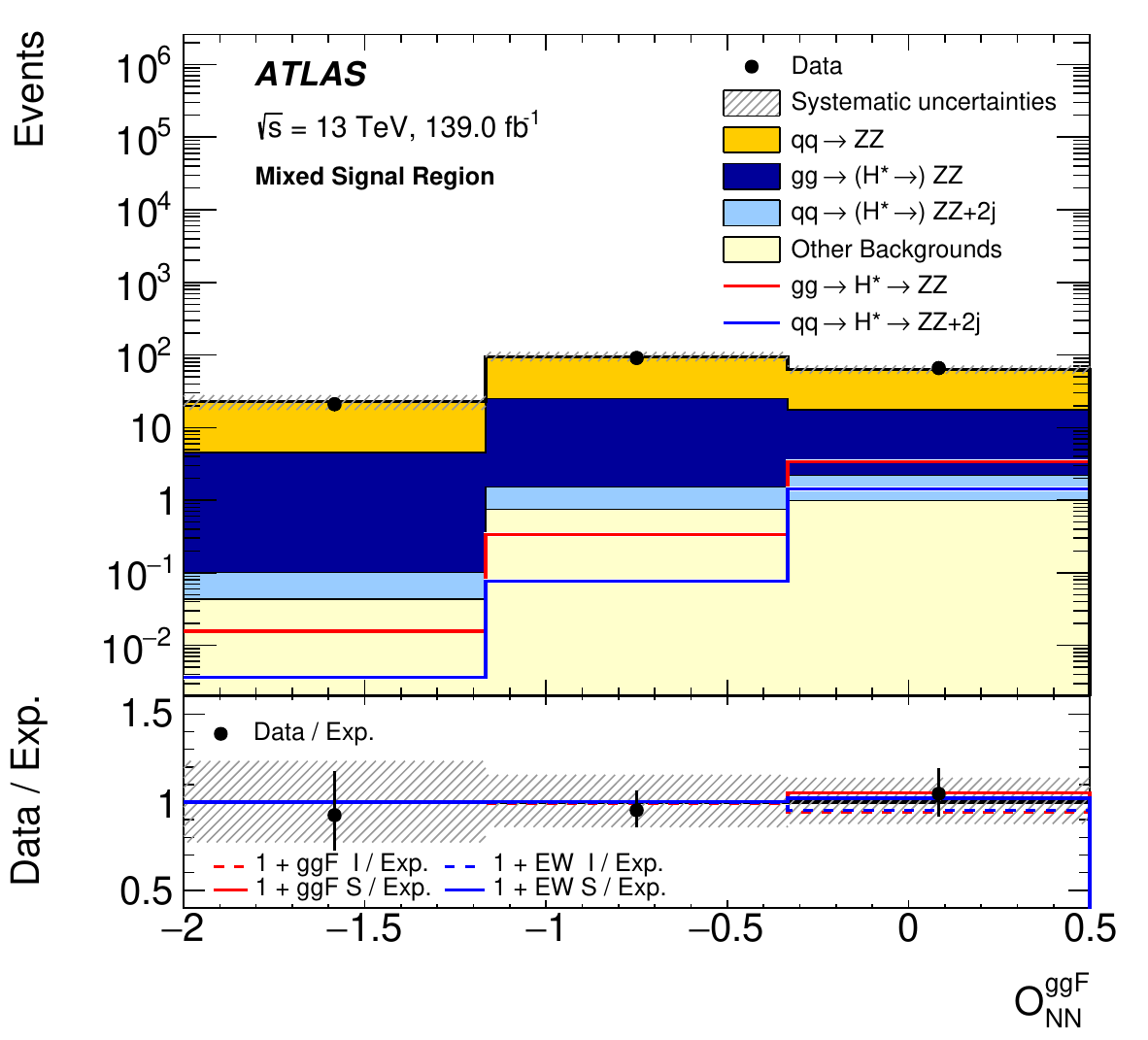}
}\\
\subfloat[]{
\includegraphics[width=0.48\textwidth]{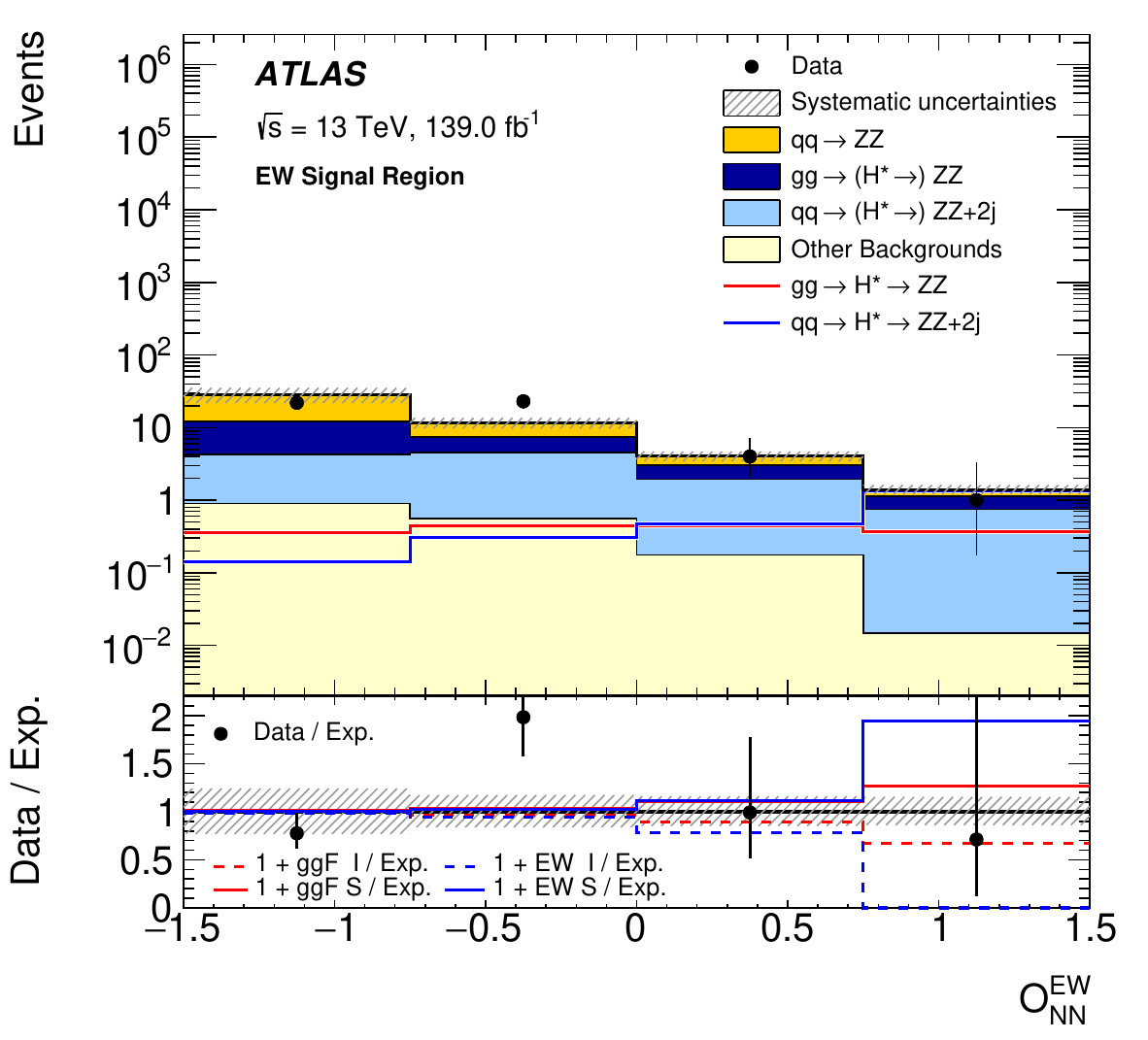}
}
\vspace{6mm}
\caption{
The observed and expected Standard Model distributions in the \fourl~channel for
(a) \ONNggF~in the ggF signal region,
(b) \ONNggF~in the mixed signal region, and (c) \ONNVBF~in the EW signal region.
The observed data are shown following the fit described in Section \ref{sec:results}, except that the fit is carried out only in the \fourl\ channel and the off-shell Higgs boson signal is fixed to the SM expectation.
The total systematic uncertainty, including all uncertainties described in Section~\ref{sec:systematics}, is shown as the hatched area including correlations between uncertainties.
The expectation includes the inclusive (signal plus background plus interference) \ggtoHtoZZ~(dark blue) and \qqtoHtoZZ~(light blue) processes, as well as the backgrounds from QCD \qqZZ~production (orange) and other processes (Z+jets, $t\bar{t}$, triboson and $t\bar{t}V$) (yellow).
The expected \ggSignal~and EW \EWSignal~signals are also shown as red and blue lines.
The first and last bins include the underflow and overflow, respectively.
The lower panel of each plot shows the ratio of data to expectation (black points) and the total systematic uncertainty (hatched area), as well as the ratio of the signal (solid lines) and the interference (dashed lines) to the expectation for ggF (red) and EW (blue) production. (For ease of display, for the last four curves one plus the ratio is plotted.)}
\label{analysis4l:PostFit_SRs_main}
\end{figure}

% End of text imported from the .//sections/analysis4l.tex input file

\FloatBarrier
\section{\zztwoltwov\ analysis}
\label{sec:analysis2l2v}

% The next lines are included from the .//sections/analysis2l2v.tex input file
The \twoltwov~final state consists of a pair of isolated leptons ($e$ or $\mu$) and large \met.
It has a larger branching fraction than the \fourl~channel, but is subject to larger background contamination.
Candidate events are preselected by requiring exactly two electrons or muons with opposite charges and \pT $>20$ \GeV.
The leading lepton must have \pT $>30$ \GeV~to surpass the trigger thresholds.
To suppress the $WZ$ background, events containing any additional lepton satisfying the loose identification criteria with $\pt>7$~\GeV~are rejected.
Requiring the dilepton invariant mass ($m_{\ell\ell}$) to be between 76 and 106 \GeV~largely reduces the contamination from the \nonresll~background, originating from $t\bar{t}$, single-top (dominated by the $Wt$ process), and $q\bar{q}\rightarrow WW$ production.
Events that satisfy this preselection are then further separated into the SRs and CRs.
To suppress the remaining background dominated by the \Zjets~ and \nonresll~ processes, further selections based on \met~ and the topology of the candidate events are applied.
Candidate events are required to have \met~$>120$ \GeV~ and \metsig~$> 10$.
The azimuthal-angle difference between the dilepton system and \vecmet, \dphizmet, must be larger than 2.5 radians, and the selected leptons must be close to each other, with the distance \drll~below 1.8.
Furthermore, the azimuthal-angle difference between any of the selected jets with $\pt > 100~\gev{}$ and \vecmet must be larger than 0.4 radians to suppress events with poorly measured jet energies.
Finally, events containing one or more $b$-jets are rejected to further suppress the \ttbar~and $Wt$ backgrounds. The selected events are then categorised into three SRs in the same way as for the \fourl~channel.

The shape and normalisation of the main background contribution from \qqZZ~production are estimated from simulation, while in the combined result the overall normalisation factors are constrained by the $ZZ$ CRs defined in the \fourl~channel as described in the previous section.
As is shown later in Table \ref{Results:normFactors_fitted}, the measured normalisations generally agree fairly well with the simulated ones, so this is a safe approach for developing the analysis in this final state.

To estimate the background from $WZ$ production, control regions enriched in $WZ$ events, with a purity of over 90\%, are defined using the full event selection given above, except that the presence of a third lepton with $\pt>20~\gev{}$ is required. %
Several further selections such as \metsig~$> 3$, a $b$-jets veto, and $m_{\mathrm{T}}^{W}$~$>60~\gev{}$,
where $m_{\mathrm{T}}^{W}$ is constructed from the third lepton's transverse momentum and the \vecmet{} vector\footnote{$m_{\mathrm{T}}^{W} \equiv \sqrt{2 \pT^{\ell} \met[1-\cos\Delta\phi({\vec{p}_{\mathrm{T}}^{\ell}},
\vec{E}_{\mathrm{T}}^{\mathrm{miss}})]}$},
are applied to suppress non-$WZ$ contributions.
Three separate $WZ$ CRs are defined according to the number of jets (zero, one, and $\geq2$ jets), and the CRs are included in the statistical fit to separately constrain the normalisation of the $WZ$ background in each $N_{jets}$ bin.
The shapes of the kinematic distributions are estimated from simulation.

To estimate the \nonresll~background, arising from $qq\rightarrow WW$, $\ttbar$, and single-top production,
a control region dominated by the \nonresll~processes (with a purity of about 95\%) is defined with all the event selection criteria
except that the final state is required to contain an opposite-sign $e\mu$ pair.
The \nonresll~contribution with the $ee$ ($\mu\mu$) pair is quite similar to that with the $e\mu$ pair, and the difference in lepton reconstruction is taken into account in the simulation.
This CR is then used to constrain the total normalisation of the \nonresll~background in all three SRs, and the kinematic shapes are modelled with simulation.
The \Zjets{} background contribution is estimated from simulation and
constrained by a normalisation factor derived in a control region enriched in
\Zjets~events. The control region is defined with all event selection criteria except that \metsig~ is required to be less than 9, and no requirements on the azimuthal angle difference between jets with $\pt > 100~\gev{}$ and \vecmet are made.
The resulting control region is about 73\% pure.
The kinematic distributions for the \Zjets~background are modelled with simulation.
The CRs for the \nonresll~and the \Zjets~backgrounds are not further divided to match the categorisation of SRs depending on jet multiplicity, due to insufficient events in the data.
Finally, minor backgrounds from the $VVV$ and $t\bar{t}V$ processes are estimated from simulation.

The distributions of the final observable \mtzz~in the \twoltwov~channel, as defined in Eq.~\ref{Intro:mTZZ}, are presented in Figure~\ref{analysis2l2v:PostFit_SRs}.
The data is shown after the simultaneous fit described in Section~\ref{sec:results}, except that the fit is carried out only in the \twoltwov\ channel and the value of $\muOFF$ is set equal to one.
All three SRs are shown together with the total systematic uncertainty from the sources described in Section~\ref{sec:systematics}.

\begin{figure}[h!]
\centering
\subfloat[]{
\includegraphics[width=0.45\textwidth]{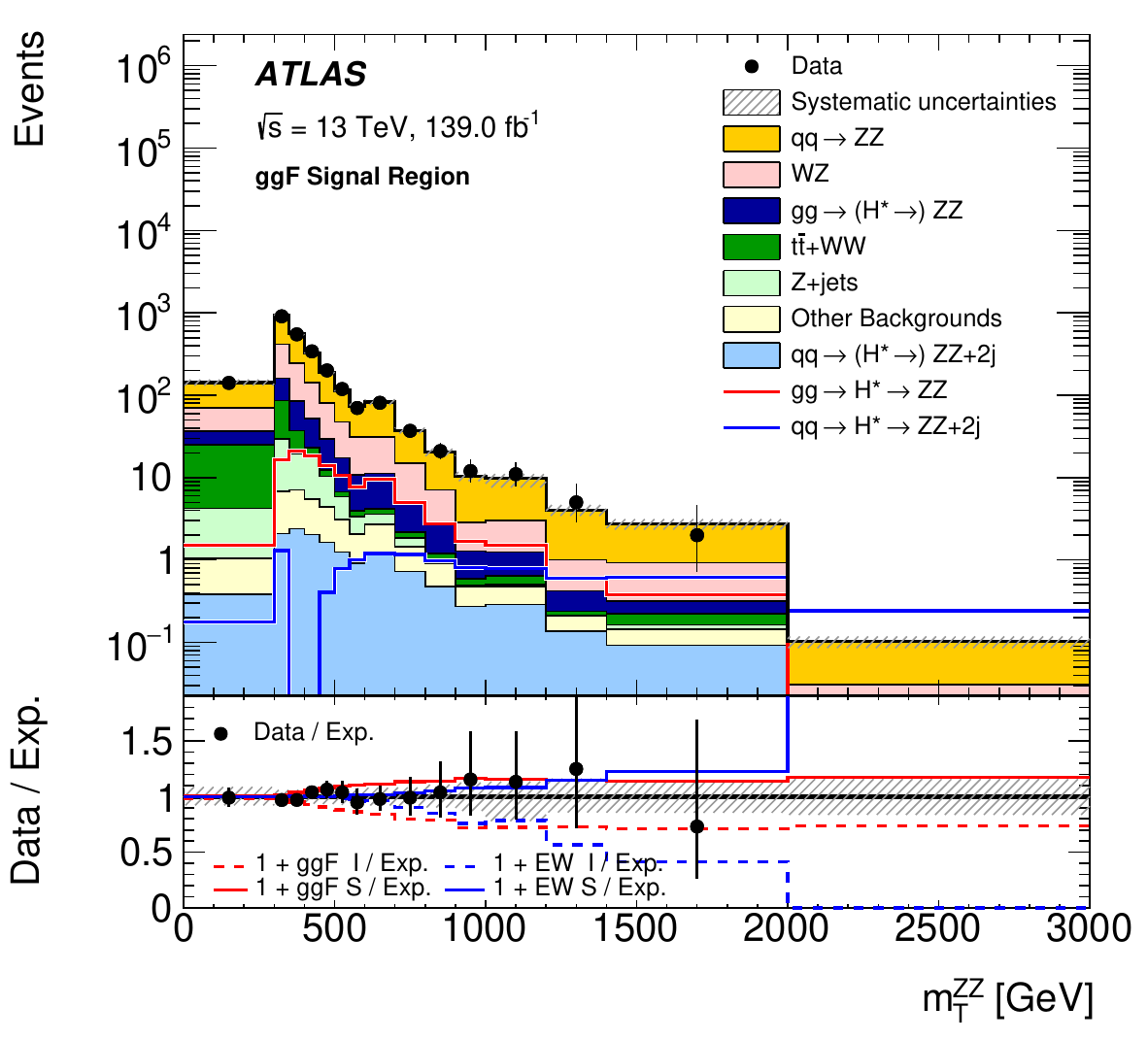}
}
\subfloat[]{
\includegraphics[width=0.45\textwidth]{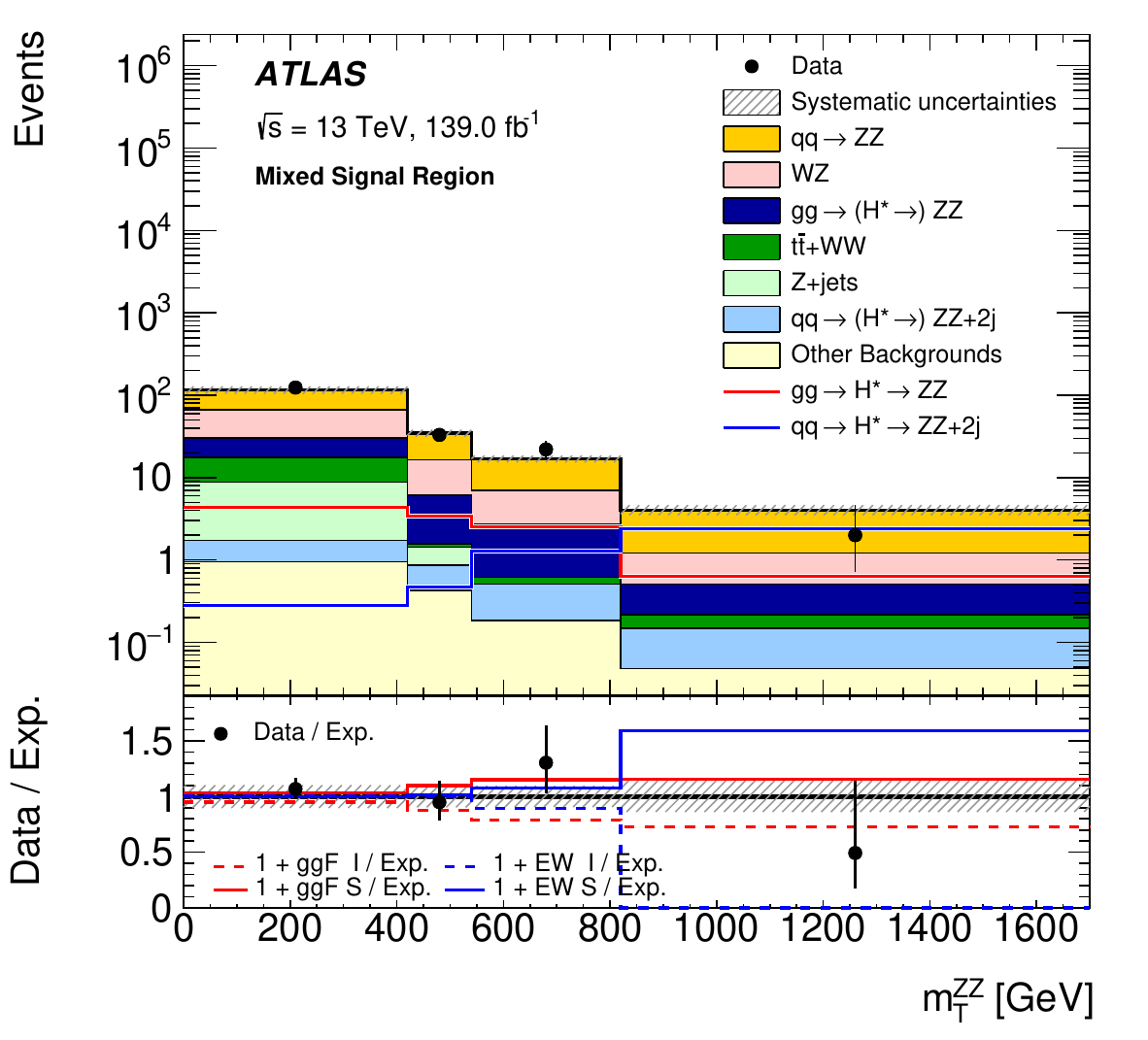}
}\\
\subfloat[]{
\includegraphics[width=0.45\textwidth]{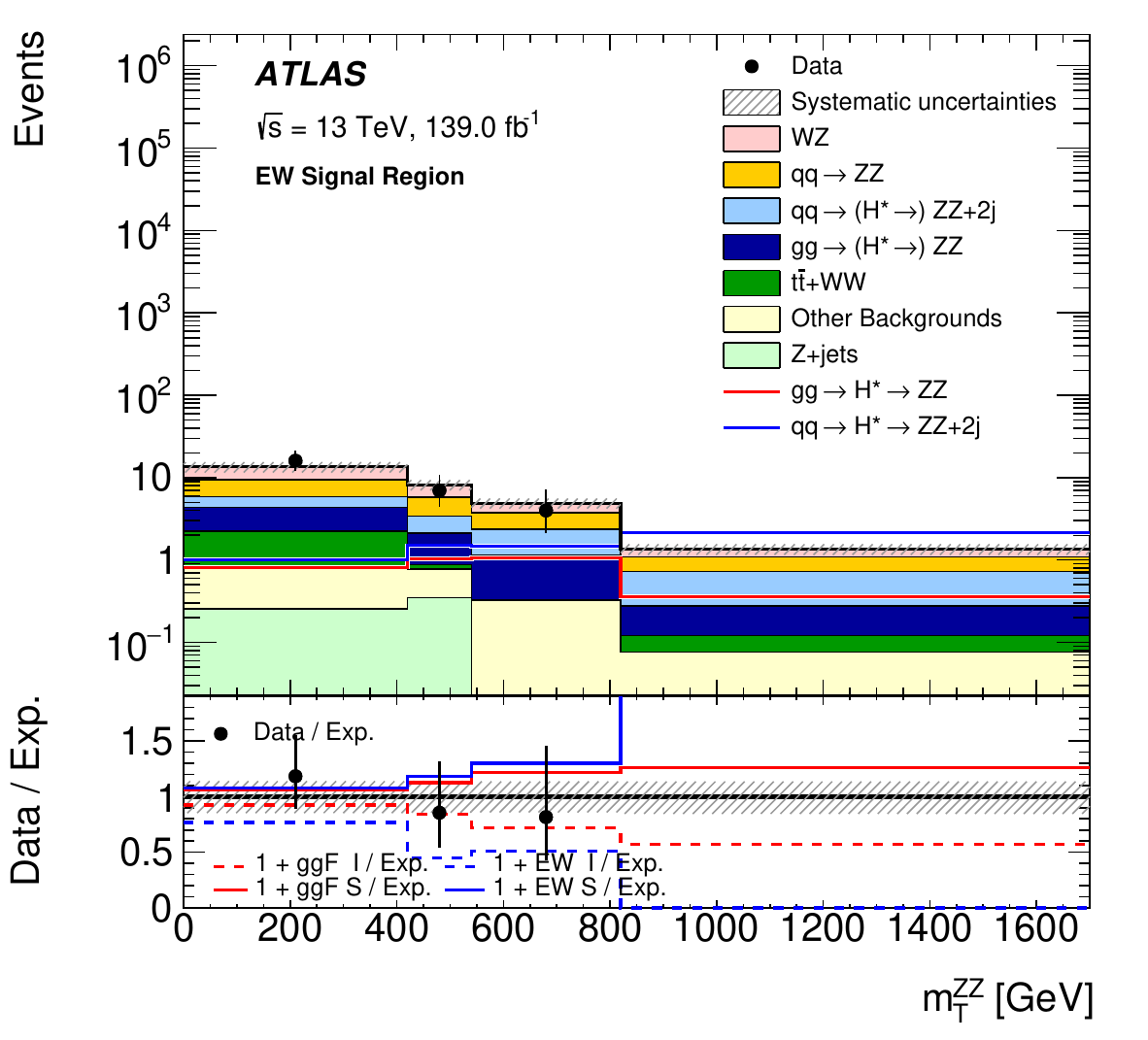}
}
\vspace{2mm}
\caption{The observed and expected Standard Model \mtzz\ distributions in the \twoltwov~channel for
(a) the ggF SR, (b) the mixed SR, and (c) the EW SR.
The data are shown after the simultaneous fit described in Section \ref{sec:results}, except that the fit is carried out only in the \twoltwov\ channel and the off-shell Higgs boson signal is fixed to the SM expectation.
The hatched area shows the total systematic uncertainty after the fit, comprising all uncertainties described in Section~\ref{sec:systematics} and including correlations between uncertainties.
The expectation includes the inclusive (signal plus background plus interference) \ggtoHtoZZ~(dark blue) and EW \qqtoHtoZZ~(light blue) production, as well as the backgrounds from QCD \qqZZ~production (orange), $WZ$ (pink), non-resonant $\ell\ell$ (dark green), $Z$+jets (light green), and other (triboson and $t\bar{t}V$) (yellow) processes.
The expected \ggSignal~and \EWSignal~signals are also shown as red and blue lines.
The last bins include the overflow.
The lower panel shows the ratio of data to expectation (black points) and the total systematic uncertainty (hatched area), as well as the ratio of the signal (solid lines) and the interference (dashed lines) to the expectation for ggF (red) and EW (blue) production.
(For ease of display, for the last four curves one plus the ratio is plotted.)}
\label{analysis2l2v:PostFit_SRs}
\end{figure}

% End of text imported from the .//sections/analysis2l2v.tex input file

\FloatBarrier
\section{Systematic uncertainties}

% The next lines are included from the .//sections/systematics.tex input file
\label{sec:systematics}

The sources of systematic uncertainty impacting the analysis of both channels can be divided into two categories:
uncertainties in the theoretical description of the signal and background processes and experimental uncertainties related to the detector response.

The largest source of systematic uncertainties arises from the theoretical modelling of the signal and background processes, including those related to the production of jets associated with the Higgs boson. Experimental uncertainties related to the reconstruction of jets are also prominent while other experimental uncertainties are generally small.
To help understand the impact of the leading uncertainties, their relative size before the statistical fit for a specific process is provided in this section, with the largest uncertainties for the main processes in the signal and control regions summarized in Table \ref{tab:systSummary}.
The impact of these uncertainties on the observed upper limits of $\muOFF$ is given in Section~\ref{sec:results}.

% The next lines are included from the .//tables/systSummary.tex input file
\begin{table*}[h!]
\caption{The dominant uncertainties in the leading processes in the signal and background regions.
Uncertainties may depend on the value of the observable: if so, a range is given in the table.
Detailed descriptions of the uncertainties are given in the text.}
\label{tab:systSummary}
\begin{center}
\renewcommand{\arraystretch}{1.2}
\renewcommand{\tabcolsep}{8pt}
\begin{tabular}{lllc}
\hline\hline
Process & Uncertainty & Final State & Value (\%) \\
\hline
\multicolumn{4}{c}{ggF Signal Region} \\
\hline
$\qqZZ$ & QCD Scale & $2\ell 2\nu$ & 4--40 \\
$\qqZZ$ & QCD Scale & $4\ell$ & 21--28 \\
$\qqZZ$ & HOEW & $4\ell$ & 1--7 \\
$\qqZZ$ & HOEW & $2\ell 2\nu$ & 2--20 \\
$\qqZZ$ & Parton Shower & $2\ell 2\nu$ & 1--67 \\
$\qqZZjj$ & Parton Shower & $2\ell 2\nu$ & 1--33 \\
$\qqZZjj$ & Parton Shower & $4\ell$ & 2--10 \\
$gg\rightarrow H^{*}\rightarrow ZZ$ & Parton Shower & $4\ell$ & 27 \\
$gg\rightarrow H^{*}\rightarrow ZZ$ & Parton Shower & $2\ell 2\nu$ & 8--45 \\
$gg\rightarrow ZZ$ & Parton Shower & $4\ell$ & 38 \\
$gg\rightarrow ZZ$ & Parton Shower & $2\ell 2\nu$ & 6--43 \\
$WZ+0j$ & QCD Scale & $2\ell 2\nu$ & 1--54 \\
\hline
\multicolumn{4}{c}{1-jet Signal Region} \\
\hline
$gg\rightarrow H^{*}\rightarrow ZZ$ & Parton Shower & $4\ell$ & 27 \\
$gg\rightarrow H^{*}\rightarrow ZZ$ & QCD Scale & $2\ell 2\nu$ & 13--18 \\
$gg\rightarrow ZZ$ & Parton Shower & $4\ell$ & 38 \\
$gg\rightarrow ZZ$ & QCD Scale & $2\ell 2\nu$ & 18--20 \\
$\qqZZjj$ & QCD Scale & $2\ell 2\nu$ & 7--18 \\
$\qqZZjj$ & QCD Scale & $4\ell$ & 3--10 \\
\hline
\multicolumn{4}{c}{2-jet Signal Region} \\
\hline
$\qqZZ$ & QCD Scale & $4\ell$ & 18--26 \\
$\qqZZ$ & QCD Scale & $2\ell 2\nu$ & 8--32 \\
$gg\rightarrow H^{*}\rightarrow ZZ$ & Parton Shower & $4\ell$ & 27 \\
$gg\rightarrow H^{*}\rightarrow ZZ$ & QCD Scale & $2\ell 2\nu$ & 14-18 \\
$gg\rightarrow ZZ$ & Parton Shower & $4\ell$ & 38 \\
$gg\rightarrow ZZ$ & QCD Scale & $2\ell 2\nu$ & 18--20 \\
$WZ+2j$ & QCD Scale & $2\ell 2\nu$ & 20--22 \\
$\qqZZjj$ & QCD Scale & $4\ell$ & 8--14 \\
$\qqZZjj$ & QCD Scale & $2\ell 2\nu$ & 8--16 \\
\hline
\multicolumn{4}{c}{$qq\rightarrow ZZ$ Control Regions} \\
\hline
$\qqZZ$ & QCD Scale & $4\ell$ & 26 \\
\hline
\multicolumn{4}{c}{Three-lepton Control Regions} \\
\hline
$WZ+2j$ & QCD Scale & $2\ell 2\nu$ & 28 \\
\hline
\hline
\end{tabular}
\end{center}
\end{table*}

% End of text imported from the .//tables/systSummary.tex input file

%
%
%

The theoretical uncertainties arise from the choice of PDF, from the missing higher-order corrections in both QCD and EW perturbative calculations, and from the modelling of the parton shower.

The PDF uncertainties are evaluated using the NNPDF prescription with MC replicas. The PDF covariance matrix between each channel of the analysis is estimated from the 100 replicas from the NNPDF3.0 NNLO set. Only the principal component of the covariance matrix has a non-negligible impact on the yields and it is used as a representation of the PDF uncertainty including its bin-by-bin correlations. The uncertainties due to missing higher-order QCD corrections are estimated by varying the renormalisation and factorisation scales independently, ranging from a factor of one-half to two (excluding the cases in which one scale is varied down by one-half and the other up by two).
For the \qqZZ~background, the uncertainty is evaluated independently in bins of $N_\textrm{jets}$.
For the gluon-induced processes, including the signal, the $gg\rightarrow ZZ$ background, and their interference, the missing higher-order uncertainties are evaluated by their impact on the respective NLO $K$-factors \cite{deFlorian:2016spz}. The uncertainties are increased in the kinematic regions of the SRs where the $K$-factor calculations are less precise due to missing effects from on-shell top quarks and high-\pt~jets~\cite{HIGG-2017-06}: the uncertainty is doubled in the phase space containing a jet with \pT $>$ 150 \GeV~ and increased by 50\% for \mzz~around twice the top-quark mass.
In both the \fourl~and \twoltwov~channels, the uncertainty from missing higher order corrections in the \qqZZ~background is one of the largest uncertainties, ranging from a few percent up to 40\% depending on jet multiplicity and observable bin.
In both channels, the same uncertainty in the gluon--gluon processes ranges from 10\% to 20\%.

The uncertainties due to missing higher-order EW corrections (HOEW) are considered for the main \qqZZ~background and handled differently in the two channels.
For the \twoltwov\ channel, the difference in the NLO EW correction between the multiplicative and additive methods as a function of \mzz\ is assigned as the uncertainty~\cite{Kallweit:2017khh}.
This uncertainty ranges from 1\% to at most 20\% of the cross-section depending on the bins of observables.
For the \fourl~channel, the NLO EW corrections are calculated on top of the LO QCD cross-section. Therefore, a specific prescription~\cite{Gieseke:2014gka}, also applied in Ref.~\cite{HIGG-2017-06}, is used to derive the uncertainty to account for missing NLO QCD+EW diagrams.
A study of the compatibility between the two methods in the \fourl\ channel shows that the central values agree to within a few percent while the uncertainties have a similar size.

The uncertainties due to the modelling of the parton shower and hadronisation play an important role in this search, as jet multiplicity, a key variable used to define both the SRs and the CRs, is particularly sensitive to the modelling of matrix elements, parton showering, and the merging and matching between the two.
The parton shower (PS) uncertainties are evaluated by varying resummation and matching scales for the processes simulated with the \SHERPA~generator.
These variations for \SHERPA~samples are expected to further account for the shape uncertainties relating to missing high-order QCD effects beyond those from the usual QCD scale variations, i.e. migrations between jet bins.
For those processes simulated with the \PYTHIA~shower program, the uncertainty is assessed by varying the \PYTHIA~configurations, such as the parameter values of the A14 tune, the multi-parton models and the final-state radiation models.
For ggF production, the PS uncertainties are correlated between the signal and background processes.
The uncertainties are split into shape and normalization components, with the latter being more significant.

In the \twoltwov~channel, the systematic uncertainties arising from the PS are parameterised using only the $ZZ$ transverse momentum.
In this channel, the PS uncertainty in the ggF processes is quite important: it is about 25\% in the yields and ranges from a few percent to a maximum of 15\% in the observable shapes.
For the EW processes, this uncertainty reaches 15\% in the yields and a few percent in the shapes.
The PS uncertainties for the \qqZZ~background are at percentage level for the bulk of observable bins but can reach up to 30\% in some parts of the phase space.

In the \fourl~channel, these uncertainties are parameterised using the $4\ell$ invariant mass, transverse momentum, and the kinematics of the leading jets. NNs are trained to estimate the density ratio between the nominal and the varied samples in this multi-dimensional space~\cite{Nachman:2021opi}. This novel method ensures a detailed description of the systematic uncertainty while reducing statistical fluctuations due to the interpolation provided by the differentiable NN.
In the \fourl~channel, PS uncertainties are the leading source of uncertainties for the gluon--gluon processes in the SRs, and are about 30\% and 40\% in the yields of the signal and the background. The impact of PS uncertainties in the shape of observables is found to be no more than 10\% in the \fourl~channel.
The PS uncertainties for the EW processes are less significant, ranging from a few percent to 10\%.
The PS uncertainties for the QCD \qqZZ~background are generally smaller, at percentage level for most observable bins.

Experimental systematic uncertainties are generally less important than the theoretical ones. However, uncertainties related to jet reconstruction are important in the $2\ell2\nu$ channel as mismeasurement of the jet energy can mimic \met. %
The main jet uncertainties are those in the jet energy scale (JES) and resolution (JER), which can amount to about 10\% for processes in the EW SRs. The effect of pile-up and the differences between the energy responses for jets with different hadron flavour compositions are particularly important. Uncertainties originating from the electron and muon reconstruction and selection, and from \met reconstruction are less important. The uncertainty in the Run-2 luminosity measurement is 1.7\%~\cite{ATLAS-CONF-2019-021}, obtained using the LUCID-2 detector \cite{LUCID2} for the primary luminosity measurements.

%
%
%

% End of text imported from the .//sections/systematics.tex input file

\FloatBarrier
\section{Results and interpretations}

% The next lines are included from the .//sections/results.tex input file
\label{sec:results}

The statistical model used to translate the results into constraints on the off-shell signal strength \muOFF~is based on the profile likelihood technique \cite{Cowan_2011}.
A binned likelihood function is constructed as a product of Poisson probability terms over all bins in all the SRs and CRs considered in the analysis, as introduced in Sections \ref{sec:analysis4l} and \ref{sec:analysis2l2v}.
The likelihood depends on the parameters of interest and a set of nuisance parameters $\theta$ that include the effects of systematic uncertainties and statistical uncertainties from the limited number of simulated events. They are constrained using Gaussian and Poisson terms, respectively. Different parameters of interest are used depending on the interpretation. The first interpretation explores two signal strength parameters ($\muOFFggF = \kappa^2_{g\textrm{, off-shell}}\kappa^2_{V\textrm{, off-shell}}$\ and $\muOFFVBF=\kappa^4_{V\textrm{, off-shell}}$) corresponding to the ggF- and EW-induced off-shell contributions, respectively. Here, $\kappa_{g}$ ($\kappa_{V}$) refers to the Higgs boson coupling to gluons (vector bosons) normalised to the SM prediction.
In the second interpretation, a single off-shell signal-strength parameter (\muOFF) is applied for all production modes, assuming that $\muOFFggF = \muOFFVBF = \muOFF$.

Given the sizeable interference effects, the off-shell signal cannot be treated independently of the interfering backgrounds.
The interference term, which is proportional to $\sqrt{\muOFF}$, must be taken into account when building the probability model. The expected number of events from the $gg\rightarrow (H^*\rightarrow) ZZ$ process for a given \muOFFggF, referred as $\nu^{{\textrm{ggF}}}$, can be obtained for a bin of the input distributions from the following parameterisation:

\begin{align*}
\nu^{{\textrm{ggF}}} (\muOFFggF,  \boldsymbol{\theta}) & = \muOFFggF \cdot n^{\textrm{ggF}}_\textrm{S}(\boldsymbol{\theta}) + \sqrt{\muOFFggF}\cdot (n^{\textrm{ggF}}_\textrm{SBI}(\boldsymbol{\theta})-n^{\textrm{ggF}}_\textrm{S}(\boldsymbol{\theta})-n^{\textrm{ggF}}_\textrm{B}(\boldsymbol{\theta}))+n^{\textrm{ggF}}_\textrm{B}(\boldsymbol{\theta})
\\
& = (\muOFFggF -\sqrt{\muOFFggF}) \cdot n^{\textrm{ggF}}_\textrm{S}(\boldsymbol{\theta}) + \sqrt{\muOFFggF} \cdot n^{\textrm{ggF}}_\textrm{SBI}(\boldsymbol{\theta}) + (1- \sqrt{\muOFFggF})\cdot n^{\textrm{ggF}}_\textrm{B}(\boldsymbol{\theta})
\end{align*}

where $n^{\textrm{ggF}}_\textrm{S}, n^{\textrm{ggF}}_\textrm{B}$ and $n^{\textrm{ggF}}_\textrm{SBI}$ represent the corresponding expected yields of the signal, the $gg\rightarrow ZZ$ background and the full $gg\rightarrow (H^*\rightarrow) ZZ$ process, respectively.
The expected number of events from the EW $q\bar{q}\rightarrow (H^*\rightarrow) ZZ +2j$ process for a given \muOFFVBF\ can be modelled similarly, and the parameterisation is determined by using three simulation samples: that for the full process $q\bar{q}\rightarrow (H^*\rightarrow) ZZ +2j$ with \muOFFVBF\ set to 1, that for the same process with \muOFFVBF\ set to 10, and the non-Higgs boson EW $q\bar{q}\rightarrow ZZ+2j$ background\footnote{The values of \muOFFVBF\ = 0, 1, and 10 are chosen for technical reasons involving the production of simulated samples but in principle any three numbers could be used.}.
The description in terms of a single signal component as performed for ggF production is not possible in the EW case because the requirement on high \mzz, used to ensure that the Higgs boson is off-shell, does not apply to the $t$-channel Higgs boson exchange in EW production, even though it is also off-shell (with t<0).
The parameterization described here ensures that this component scales with \muOFFVBF.

The total normalisation of the \qqZZ~background is left as a free parameter in the profile likelihood fit, separately for each jet multiplicity. Three parameters are introduced in the likelihood model to constrain the normalisation of this dominant background in both final states, a normalization factor for 0-jet events, $\mu_{qqZZ}$, and two additional parameters to represent the relative contributions of higher jet multiplicities, $\mu_{qqZZ}^{1j}$, and $\mu_{qqZZ}^{2j}$.
The expected yield of 0-jet \qqZZ~events is scaled by $\mu_{qqZZ}$, that of 1-jet events by $\mu_{qqZZ}\cdot \mu_{qqZZ}^{1j}$, and that of events with at least two jets by $\mu_{{qqZZ}} \cdot \mu_{qqZZ}^{1j} \cdot \mu_{qqZZ}^{2j}$.
These parameters are largely constrained by the three CRs defined in Section~\ref{sec:analysis4l}, especially at high jet multiplicity.

The normalisations of the $WZ$, \Zjets~and \nonresll~backgrounds are also obtained from the simultaneous fit, using the dedicated control regions described in Section~\ref{sec:analysis2l2v}.
Similarly to the \qqZZ~background, events from the $WZ$ process are treated separately for each jet multiplicity. Five additional free parameters, $\mu_{3\ell}$, $\mu_{3\ell}^{1j}$, $\mu_{3\ell}^{2j}$, $\mu_{Zj}$, and $\mu_{e\mu}$, are therefore introduced in the likelihood model specifically for the \twoltwov~ channel and for its combination with the \fourl~channel.

The likelihood function for the combination of both channels is built as a product of the likelihoods of the individual channels. Theoretical and experimental uncertainties with common sources are treated as correlated between the two channels. The NLO EW uncertainty is uncorrelated between the two channels, due to the different schemes used to derive the uncertainties.
The hypothesis of systematic uncertainty correlation between the $4\ell$ and $2\ell2\nu$ channels is tested for the dominant sources of uncertainties, including the PS uncertainties that use models with different complexity in the two channels, and the NLO EW uncertainty. The difference in the result when using different correlation hypotheses is found to be negligible.

The \mfourl\ distribution for the \fourl~channel and the \mtzz\ distribution for the \twoltwov~channel are shown in Figure~\ref{Results:VRs} after the full fit to data with $\muOFF=1$. The total systematic uncertainty from the sources described in Section~\ref{sec:systematics} are shown in the figure. The distributions of the NN observables used in the \fourl~channel are shown in Figure~\ref{analysis4l:PostFit_SRs_main}.

\begin{figure}[h!]
\centering
\subfloat[]{\includegraphics[width=0.46\linewidth]{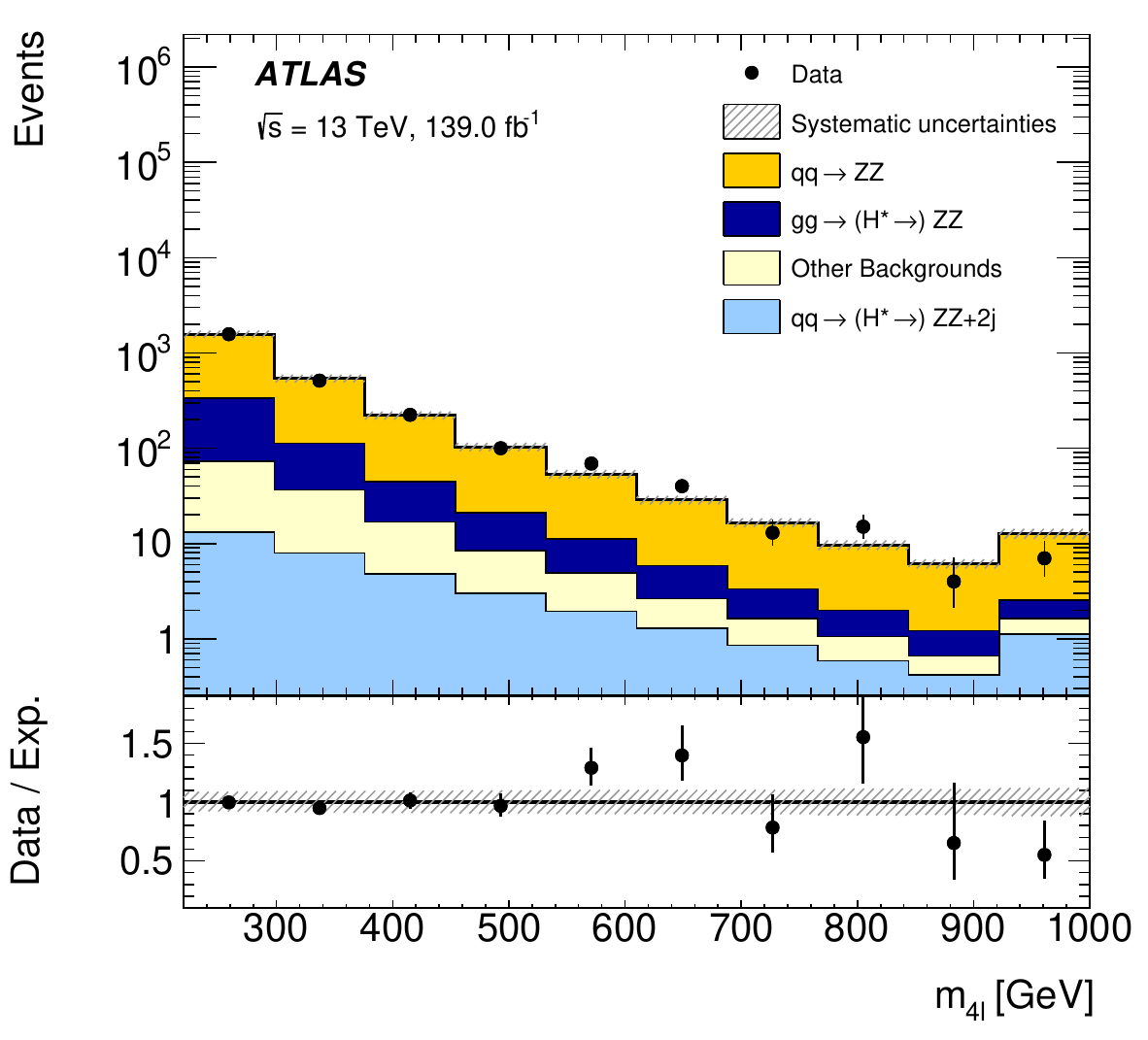}}%
\subfloat[]{\includegraphics[width=0.46\linewidth]{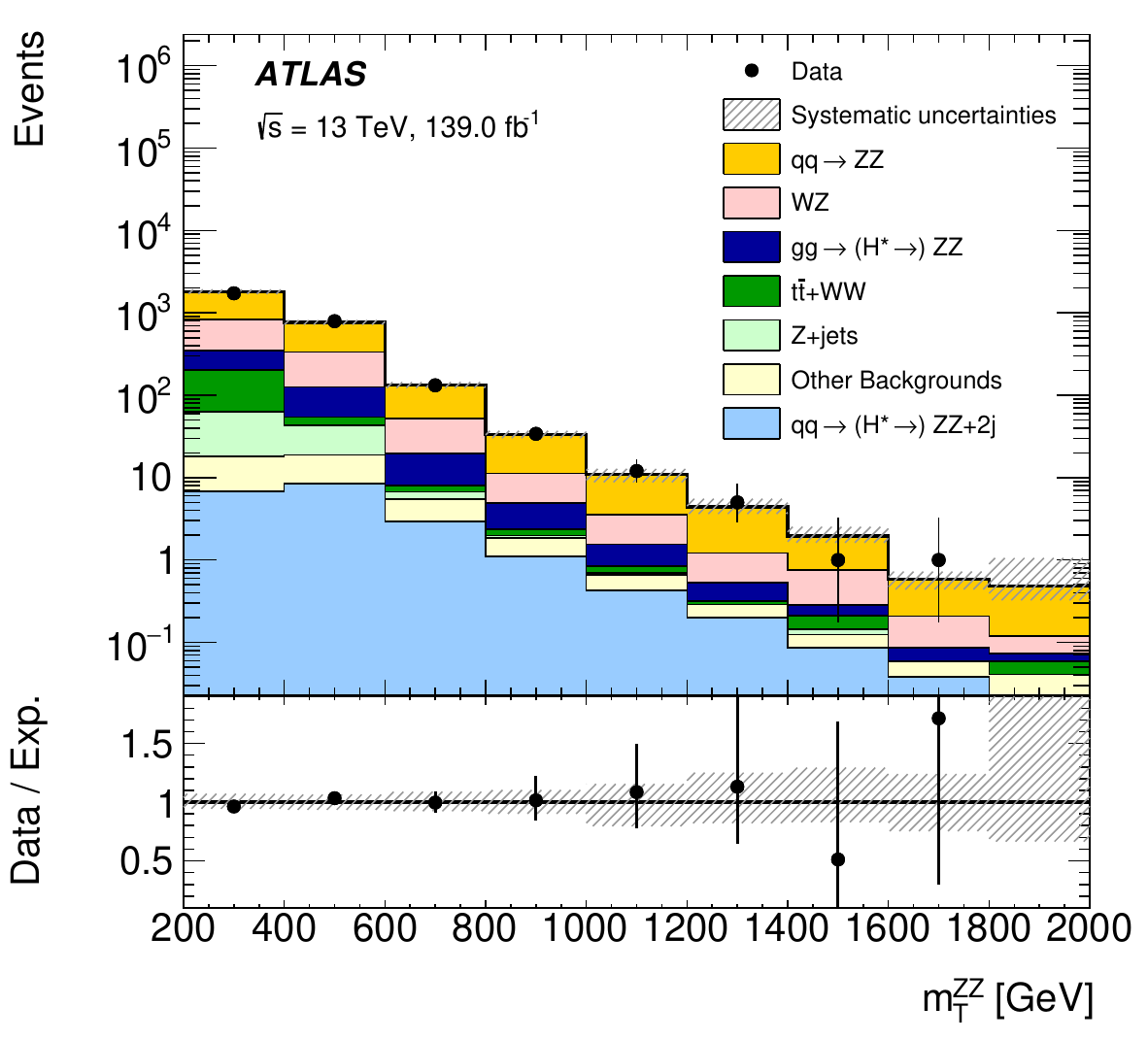}}%
\caption{Comparisons between data and the SM prediction for the (a) \mfourl\ and (b) \mtzz\ distributions in the inclusive off-shell signal regions in the \zzfourl\ and \zztwoltwov\ channels, respectively. The scenario with the off-shell signal strength equal to one is considered in the fit. The hatched area represents the total systematic uncertainty. The last bin in both figures contains the overflow.}
\label{Results:VRs}
\end{figure}

The expected numbers of events in the SRs after the maximum-likelihood fit to the data performed in all SRs and CRs, together with the corresponding observed yields, are shown in Tables~\ref{Results:SR_yields_4l} and~\ref{Results:SR_yields_2l2v} for the \zzfourl\ and \zztwoltwov\ channels, respectively. The fitted background normalisation factors together with their total uncertainties are summarized in Table \ref{Results:normFactors_fitted}.

% The next lines are included from the .//tables/yields4l_mu1_SI_NEW.tex input file
%

\begin{table*}[h!]
\caption{The observed and expected yields together with their uncertainties, for the ggF- and EW-enriched categories in the \fourl\ channel.
The results are obtained after the simultaneous fit to both the \fourl~and \twoltwov~channels with $\muOFF = 1$.
The first row represents the inclusive $ZZ$ process from $gg$ production, including the signal, background, and interference components.
The signal and background components are shown separately in rows 2--3: they do not add up to match the inclusive yield due to the presence of negative interference.
The other backgrounds include contributions from $t\bar{t}V$ and $VVV$ processes. The uncertainties in the expected number of events include the statistical and systematic uncertainties. The uncertainties in the $qq\rightarrow ZZ$ background are quoted as the sum in quadrature of all three jet multiplicity contributions for purposes of illustration.}
\label{Results:SR_yields_4l}
\newcolumntype{e}[1]{D{+}{\,\pm\,}{#1}}
\begin{center}
\renewcommand{\arraystretch}{1.2}
\renewcommand{\tabcolsep}{8pt}
\begin{tabular}{le{6.6}e{6.6}e{6.6}}
\hline\hline
Process & \multicolumn{1}{c}{ggF SR} & \multicolumn{1}{c}{Mixed SR} & \multicolumn{1}{c}{EW SR} \\
\hline%
$gg\rightarrow (H^*\rightarrow) ZZ$ & 341 + 117 & 42.5 + 14.9 & 11.8 + 4.3\\
\qquad $gg\rightarrow H^*\rightarrow ZZ$ & 32.6 + 9.07 & 3.68 + 1.03 & 1.58 + 0.47\\
\qquad $gg\rightarrow ZZ$ & 345 + 119 & 43.0 + 15.2 & 11.9 + 4.4\\
$ q\bar{q}\rightarrow (H^*\rightarrow) ZZ+2j$ & 23.2 + 1.0 & 2.03 + 0.16 & 9.89 + 0.96\\
\hline
$\qqZZ$  & 1878 + 151 & 135 + 23 & 22.0 + 8.3\\

Other backgrounds & 50.6 + 2.5 & 1.79 + 0.16 & 1.65 + 0.16\\
\hline
Total expected (SM) & 2293 + 209 & 181 + 29 & 45.3 + 10.0\\
Observed & \multicolumn{1}{c}{2327} & \multicolumn{1}{c}{178} & \multicolumn{1}{c}{50} \\ \hline

\hline\hline\noalign{\smallskip}
\end{tabular}
\end{center}
\vspace{4mm}
\vspace{-4mm}
\end{table*}

%

% End of text imported from the .//tables/yields4l_mu1_SI_NEW.tex input file

% The next lines are included from the .//tables/yields2l2v_mu1_SI_NEW.tex input file
\begin{table*}[h!]
\caption{The observed and expected yields together with their uncertainties, for the ggF- and EW-enriched categories in the \twoltwov\ channel.
The results are obtained after the simultaneous fit to both the \fourl~and \twoltwov~channels with $\muOFF = 1$.
The first row represents the inclusive $ZZ$ process from $gg$ production, including the signal, background, and interference components.
The signal and background components are shown separately in rows 2--3: they do not add up to match the inclusive yield due to the presence of negative interference.
The other backgrounds include contributions from $VVV$, $W$+jets and top quark processes other than pair production. The uncertainties in the expected number of events include the statistical and systematic uncertainties. The uncertainties in the $qq\rightarrow ZZ$ and $WZ$ backgrounds are quoted as the sum in quadrature of all three jet multiplicity contributions for purposes of illustration.}
\label{Results:SR_yields_2l2v}
\newcolumntype{e}[1]{D{+}{\,\pm\,}{#1}}
\begin{center}
\renewcommand{\arraystretch}{1.2}
\renewcommand{\tabcolsep}{8pt}
\begin{tabular}{le{6.6}e{6.6}e{6.6}}
\hline\hline
Process & \multicolumn{1}{c}{ggF SR} & \multicolumn{1}{c}{Mixed SR} & \multicolumn{1}{c}{EW SR} \\
\hline%
$gg\rightarrow (H^*\rightarrow) ZZ$ & 210 + 53 & 19.7 + 4.9 & 4.29 + 1.10\\
\qquad $gg\rightarrow H^*\rightarrow ZZ$ & 111 + 26 & 10.9 + 2.5 & 3.26 + 0.82\\
\qquad $gg\rightarrow ZZ$ & 251 + 66 & 23.4 + 6.2 & 5.31 + 1.46\\
$ q\bar{q}\rightarrow (H^*\rightarrow) ZZ+2j$ & 14.0 + 3.0 & 1.63 + 0.17 & 4.46 + 0.50\\
\hline
$\qqZZ$ & 1422 + 112 & 80.4 + 11.9 & 7.74 + 2.99\\
$WZ$ & 678 + 54 & 51.9 + 6.9 & 7.89 + 2.50\\
$Z$+jets            & 62.3 + 24.3 & 7.51 + 6.94 & 0.62 + 0.54\\
\Nonresll             & 106 + 39 & 9.17 + 2.73 & 1.55 + 0.42\\
Other backgrounds   & 22.6 + 5.2 & 1.62 + 0.25 & 1.40 + 0.10\\
\hline
Total expected (SM) & 2515 + 165 & 172 + 17 & 28.0 + 4.1\\
Observed & \multicolumn{1}{c}{2496} & \multicolumn{1}{c}{181} & \multicolumn{1}{c}{27} \\ \hline

\hline\hline\noalign{\smallskip}
\end{tabular}
\end{center}
\vspace{4mm}
\vspace{-4mm}
\end{table*}

% End of text imported from the .//tables/yields2l2v_mu1_SI_NEW.tex input file

% The next lines are included from the .//tables/normFactors.tex input file
\begin{table*}[h!]
\caption{The fitted normalization factors for the dominant $q\bar{q}\rightarrow ZZ$ background as well as the $WZ$, $Z$+jets and non-resonant-$\ell\ell$ type backgrounds.}
\label{Results:normFactors_fitted}
\begin{center}
\renewcommand{\arraystretch}{1.2}
\renewcommand{\tabcolsep}{8pt}
\begin{tabular}{lc}
\hline\hline
Normalization factor & Fitted value \\
\hline%
$\mu_{\textrm{qqZZ}}$ & $1.11 \pm  0.07$\\
$\mu_\textrm{qqZZ}^{1j}$ & $0.90 \pm  0.10$\\
$\mu_\textrm{qqZZ}^{2j}$ & $0.88 \pm  0.26$\\ \hline
$\mu_{3\ell}$ & $1.06 \pm  0.03$\\
$\mu_{3\ell}^{1j}$ & $0.92 \pm  0.10$\\
$\mu_{3\ell}^{2j}$ & $0.75 \pm  0.19$\\
$\mu_{\textrm{Zj}}$ & $0.90 \pm  0.19$\\
$\mu_{e\mu}$ & $1.08 \pm  0.09$\\
\hline\hline\noalign{\smallskip}
\end{tabular}
\end{center}
\vspace{4mm}
\vspace{-4mm}
\end{table*}

% End of text imported from the .//tables/normFactors.tex input file

% The next lines are included from the .//tables/syst_impact_short.tex input file
\begin{table*}[h!]
\caption{
The impact of most important systematic uncertainties on the observed upper value of \muOFF~for which $-2\ln\lambda = 4$, obtained by the combined fit. This value corresponds to the two standard deviation upper limit of \muOFF~with the asymptotic method. The first column denotes the systematic uncertainty that was excluded from the fit. The last row gives the nominal upper limit, where all uncertainties are included. The further the upper limit is deviating from the last row value, the more important that uncertainty is.
}
\label{Results:syst_impact}
\begin{center}
\renewcommand{\arraystretch}{1.2}
\renewcommand{\tabcolsep}{8pt}
\begin{tabular}{lc}
\hline\hline
Systematic Uncertainty Fixed & \muOFF\ value at which $-2\ln\lambda(\muOFF) = 4$ \\
\hline
Parton shower uncertainty for $gg\rightarrow ZZ$ (normalisation) & 2.26 \\
Parton shower uncertainty for $gg\rightarrow ZZ$ (shape) & 2.29 \\
NLO EW uncertainty for $q\bar{q}\rightarrow ZZ$ & 2.27  \\
NLO QCD uncertainty for $gg\rightarrow ZZ$ & 2.29  \\
Parton shower uncertainty for $q\bar{q}\rightarrow ZZ$ (shape) & 2.29 \\
\hline
Jet energy scale and resolution uncertainty & 2.26  \\
\hline
None    & 2.30  \\
\hline\hline\noalign{\smallskip}
\end{tabular}
\end{center}
\vspace{4mm}
\vspace{-4mm}
\end{table*}

% End of text imported from the .//tables/syst_impact_short.tex input file

To obtain the results for a given parameter of interest, profile likelihood ratios (denoted by $\lambda$) are computed for different values of each parameter. The $-2\ln\lambda$ curve as a function of $\muOFF$ is presented in Figure~\ref{results:NLL_muoff}.
The expected curve is constructed from a fit to an Asimov dataset which is built from the SM expectation.
The expected curve is flatter than the observed due to the effect of a downward fluctuation in the data and the parabolic shape of the yield versus $\mu$ curve, which arises due to the $\sqrt{\mu}$ dependence of the interference.
In particular, for electroweak production, the expected yield is minimized at the value of $\mu=0.8$, close to the value $\mu=1$ at which the expected $-2\ln\lambda$ curve is minimized.
In this case, a downward fluctuation in the data, observed for this production mode, does not appreciably move the minimum, because a lower yield cannot reduce the best-fit value of $\mu$ below 0.8.
Instead, the lower yield makes values further from the minimum less likely, thus narrowing the profile likelihood.

Due to the quadratic parameterisation of the yield as a function of the parameter of interest, the distribution of the test statistic $-2\ln\lambda$ is slightly different from the asymptotic $\chi^{2}$ distribution predicted by Wilks’ theorem~\cite{10.1214/aoms/1177732360}.
Therefore confidence intervals on \muOFF~are built based on the Neyman construction~\cite{10.2307/91337} using the distribution of $-2\ln\lambda$ for different values of the parameter of interest. The distributions of $-2\ln\lambda$ are estimated using simulated events sampled from the likelihood model of the analysis, profiled to the best fit results from data.
The resulting confidence intervals are 5--10\% more conservative than those obtained by assuming that the asymptotic assumption is correct.
The intersection of the 1-sigma and 2-sigma curves in Figure \ref{results:NLL_muoff} with the likelihood curves allow the true 68 and 95\% confidence intervals to be estimated.
The expected uncertainty in \muOFF\, also obtained using the $1\sigma$ confidence intervals from the Neyman construction, is $\pm0.9$.
The observed value of \muOFF\ with the $1\sigma$ confidence intervals from the Neyman construction is $\muOFF=1.1^{+0.7}_{-0.6}$.
The observed (expected) 95\% confidence level (CL) upper limit on \muOFF\ is 2.4 (2.5). The background-only hypothesis (\muOFF $=0$) is rejected at an observed (expected) significance of 3.3$\sigma$ (2.2$\sigma$).
The $-2\ln\lambda=2.30$ and $-2\ln\lambda=5.99$ 2D contours for \muOFFggF~ and \muOFFVBF, which correspond to the 68\% and 95\% CL limits in the asymptotic approximation, are shown in Figure~\ref{results:NLL_2D}.

To estimate the importance of the most relevant sources of systematic uncertainty to the result, Table~\ref{Results:syst_impact} shows the value of the largest \muOFF\ for which $-2\ln\lambda=4$ when each source of uncertainty is removed one at a time.
Due to the unusual shape of the yield curve, the impact of nuisance parameters on the best-fit value of \muOFF~can be difficult to interpret.
Additionally, the correlations between the nuisance parameters, and between the nuisance parameters and the normalization parameters for the backgrounds, make it difficult to extract uncertainty components.
Table~\ref{Results:syst_impact} indicates the magnitude of the systematic uncertainties and shows their relative importance. The most important ones are the PS uncertainties, the NLO EW uncertainties, and the jet-related uncertainties.

The combination with the on-shell $H\rightarrow ZZ^*\rightarrow 4\ell$ analysis~\cite{HIGG-2018-28}, where the on-shell signal strength is measured to be $\muON=1.01\pm 0.11$, allows these results to be translated into limits on the width of the Higgs boson normalised to its SM expectation ($\muOFF/\muON=\gammaRatio$) as well as the ratio of off-shell to on-shell couplings for ggF ($R_{gg} \equiv \kappa^2_{g\textrm{, off-shell}}/\kappa^2_{g\textrm{, on-shell}}$) and EW ($R_{VV} \equiv \kappa^2_{V\textrm{, off-shell}}/\kappa^2_{V\textrm{, on-shell}}$) production.
The experimental uncertainties are correlated between the two measurements, while the theoretical uncertainties are assumed to be uncorrelated, considering that differences could exist in the structure of high-order corrections at different mass scales.
The difference in the statistical results between the correlated and uncorrelated schemes is found to be negligible.
The \gammaRatio\ interpretation assumes that the off- and on-shell coupling modifiers are the same for both ggF and EW production modes.
The $R_{gg}$ and $R_{VV}$ interpretations assume that the total width of the Higgs boson is equal to its SM prediction, and that the scattering phase also follows the SM.
Additionally, in the \rgg\ case it is assumed that the coupling scale factors associated with the on- and off-shell EW production are the same, while in the \rvv\ case the $t$-channel Higgs boson exchange process is assumed to scale in the same way as for the off-shell signal.

For the combination with the on-shell analysis, the combined likelihood is built as the product of the likelihood models for the two analyses.
The values of $-2\ln\lambda$ as a function of \gammaRatio, \rgg\ and \rvv\ are shown in Figures~\ref{results:NLL_relwidth},~\ref{results:NLL_rgg} and~\ref{results:NLL_rvv}, respectively.
The deviation of $-2\ln\lambda$ from a smooth parabolic curve in the region close to zero in Figure~\ref{results:NLL_rgg} is due to the $\sqrt{\mu}$ dependence in the yield that arises from the interference, as discussed earlier in this section, combined with a slight excess observed in the data in the \fourl\ ggH SR, which leads to a maximum near $\mu=0$.
Confidence intervals are obtained using the Neyman construction, as described above.
The corresponding measured values are the following: $\gammaRatio=1.1^{+0.7}_{-0.5}$, $\rgg=1.2^{+1.0}_{-1.2}$ and $\rvv=0.9^{+0.5}_{-0.3}$.
Multiplying the measured $\gammaRatio$ by the width of the SM Higgs boson, the measured \gammaH~is $4.4^{+3.0}_{-2.2}$ \MeV.
The total uncertainty is dominated by its statistical component.
The observed (expected) upper limit on \gammaRatio\ is 2.5 (2.6) at 95\% confidence level using the Neyman construction, and the corresponding lower limit is 0.1 (0.01).
Thus observed (expected) upper and lower limits can be placed on the total width of the Higgs boson of $0.5 (0.1) < \Gamma_H < 10.2 (10.9)$ \MeV.

\begin{figure}[h!]
\centering
\subfloat[]{\includegraphics[width=0.47\linewidth]{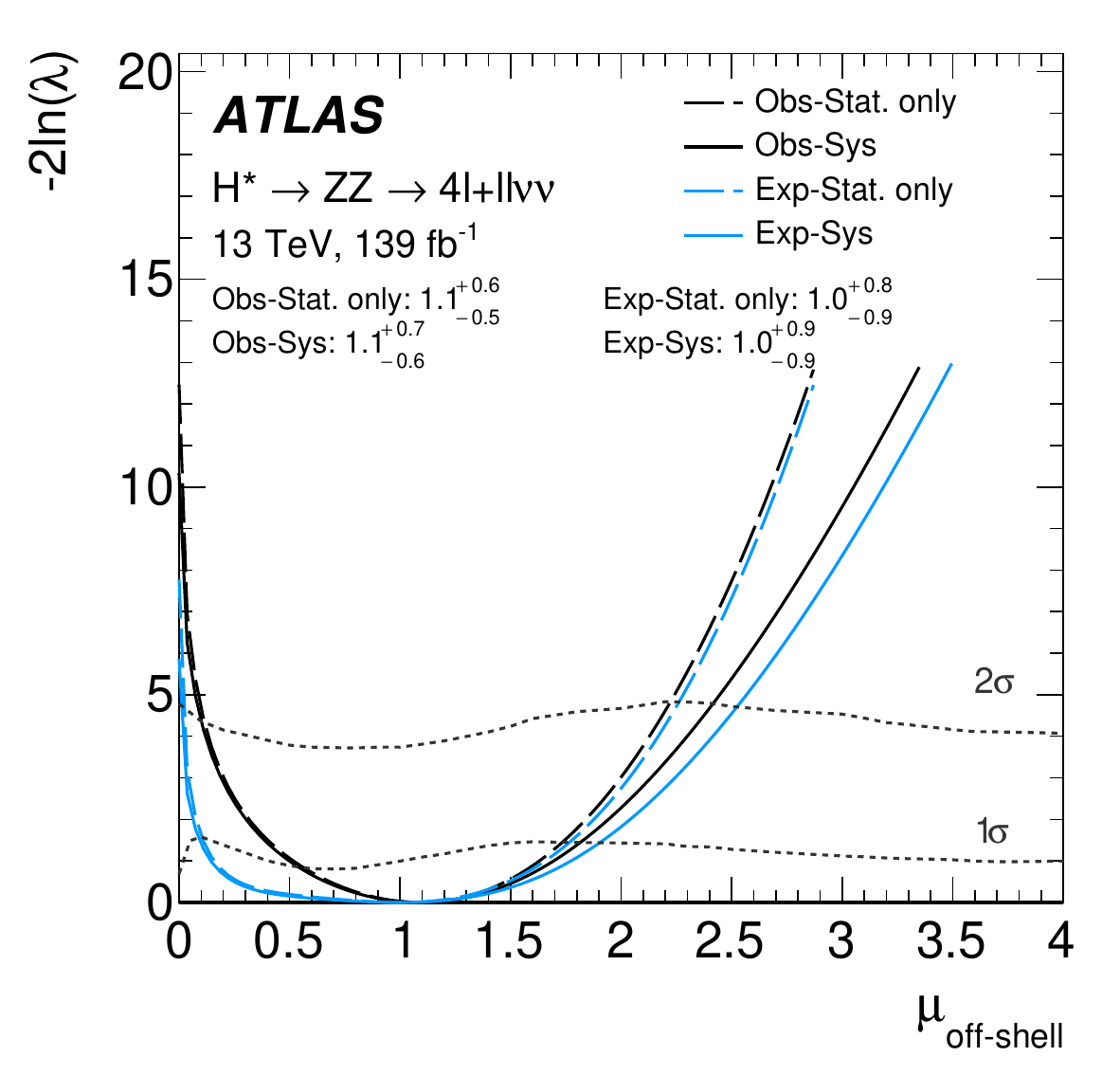}\label{results:NLL_muoff}}
\subfloat[]{\includegraphics[width=0.47\linewidth]{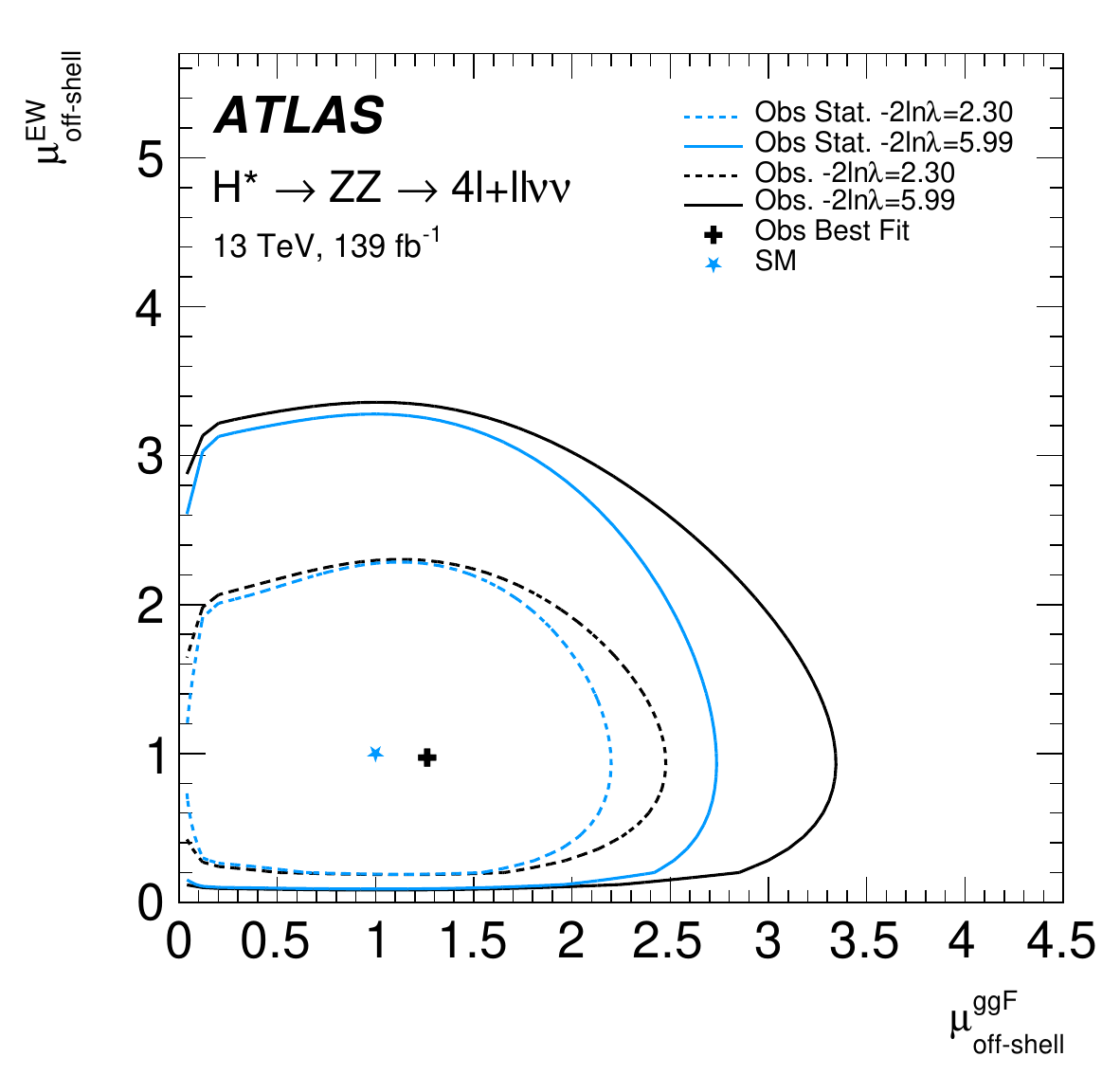}\label{results:NLL_2D}}
\caption{The likelihood profile, $-2\ln\lambda $, as a function of (a) the off-shell Higgs boson signal strength, \muOFF, for the combination of the \zzfourl\ and \zztwoltwov\ off-shell analyses, and (b) two off-shell signal strength parameters for the ggF and EW production modes, plotted in a plane (\muOFFggF,  \muOFFVBF).
In (a) the dotted curves correspond to the one and two standard deviation confidence intervals on the measurement obtained using the Newyman construction while (b) shows two-dimensional contours for \muOFFggF~ and \muOFFVBF\ corresponding to $-2\ln\lambda=2.30$ and $-2\ln\lambda=5.99$, which correspond to the 68\% and 95\% CL limits in the asymptotic approximation.}
\end{figure}

\begin{figure}[h!]
\centering
\subfloat[]{\includegraphics[width=0.46\linewidth]{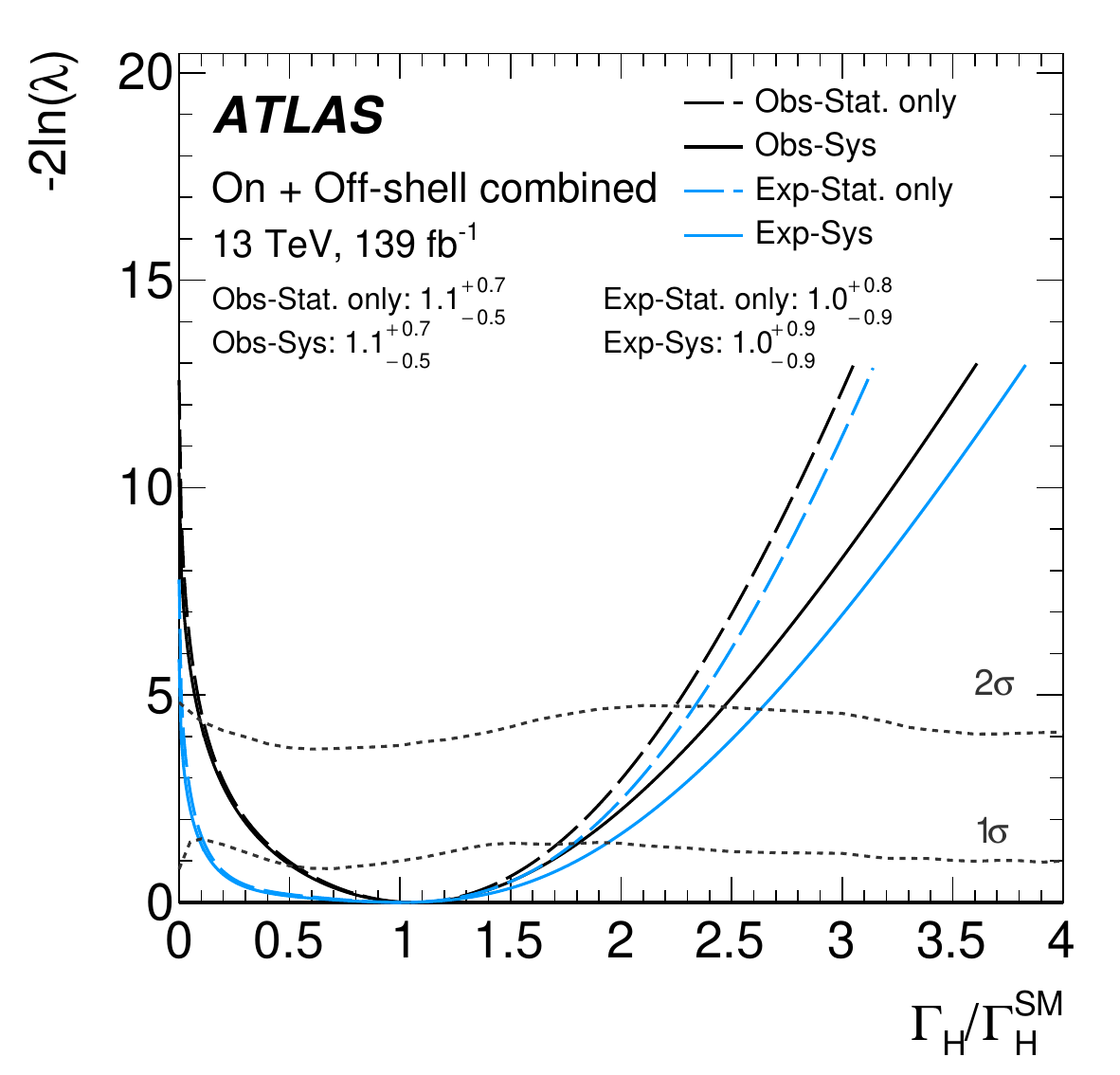}\label{results:NLL_relwidth}}\\
\subfloat[]{\includegraphics[width=0.46\linewidth]{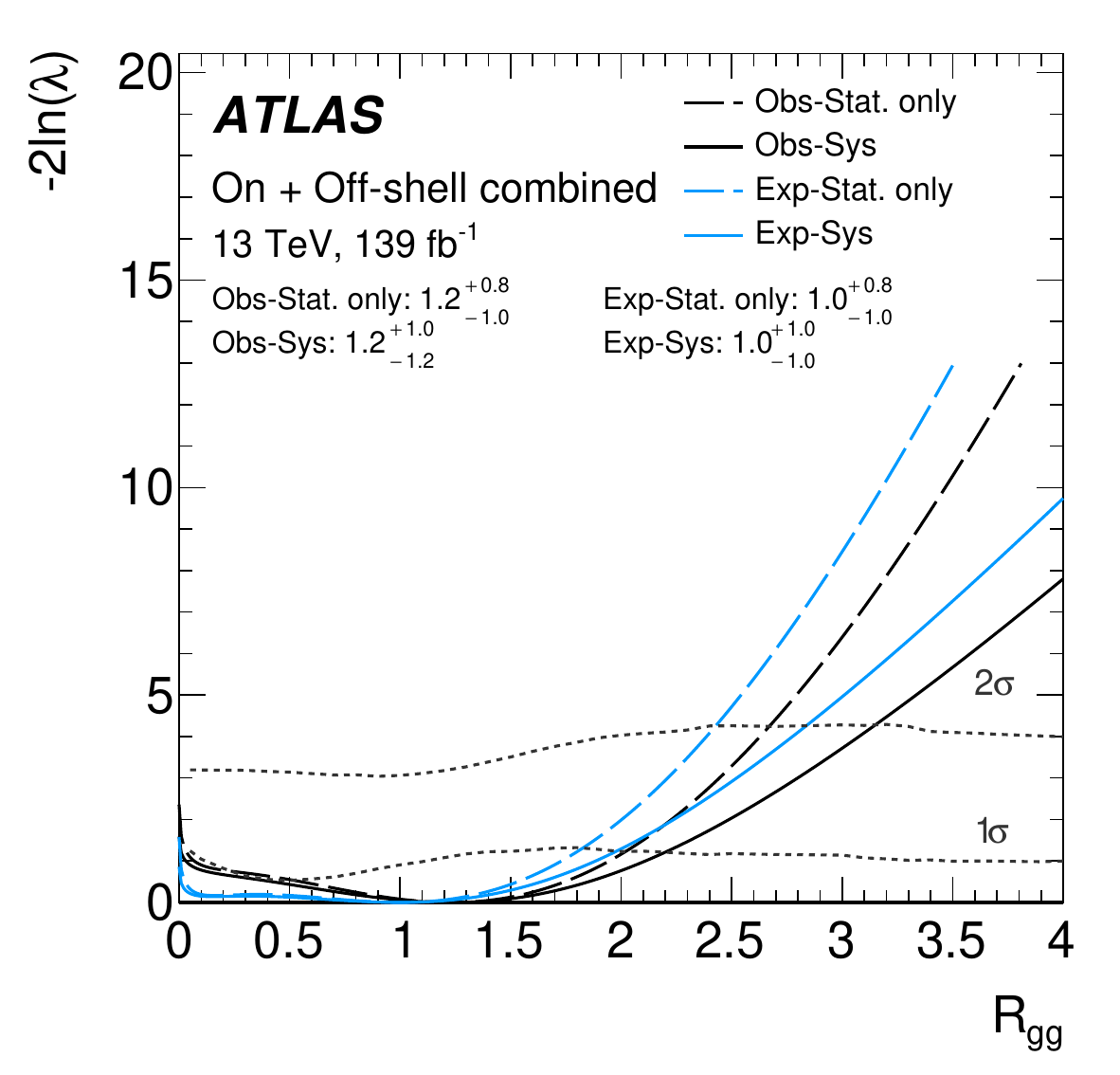}\label{results:NLL_rgg}}
\subfloat[]{\includegraphics[width=0.46\linewidth]{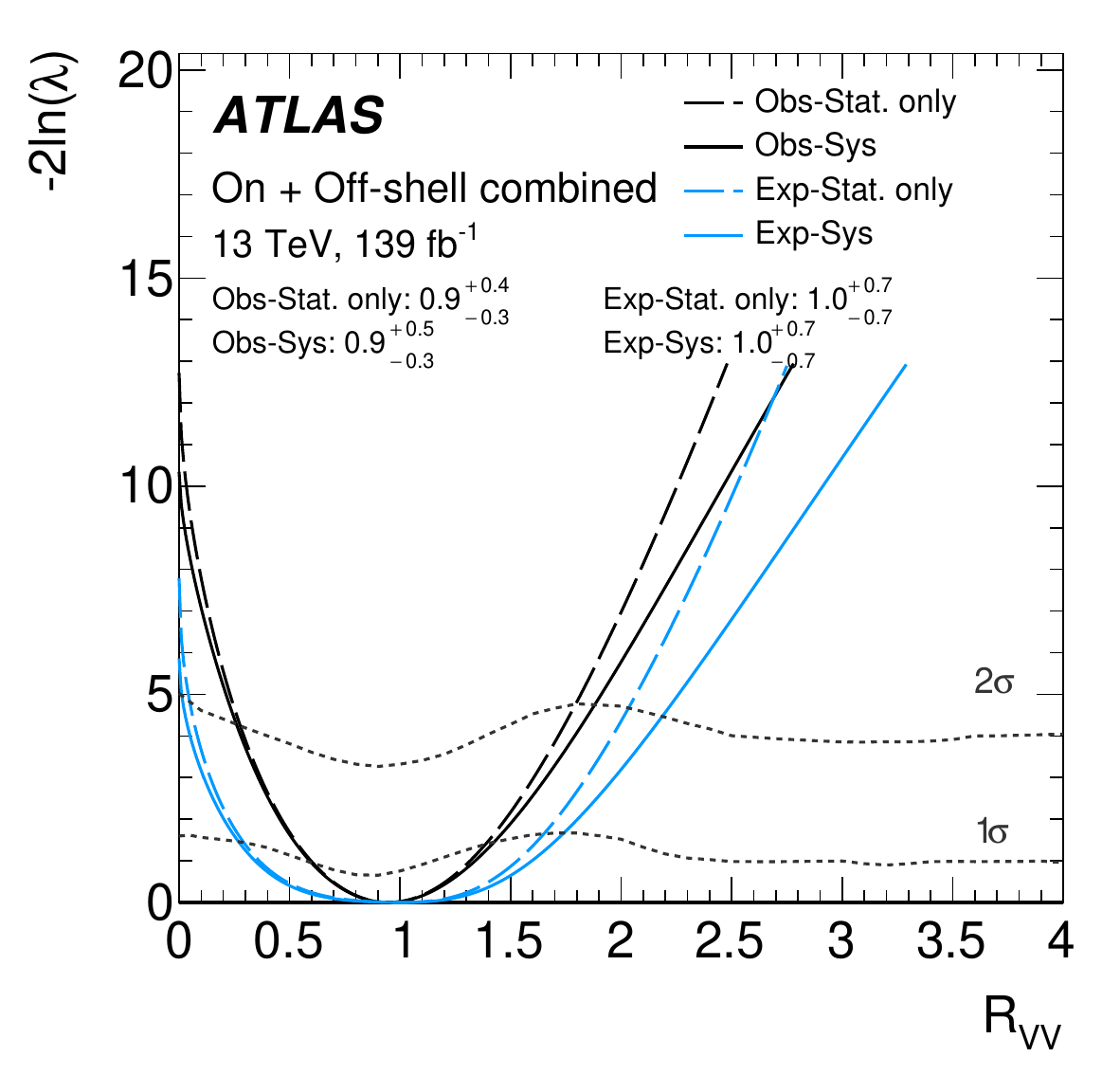}\label{results:NLL_rvv}}
\caption{The likelihood profile, $-2\ln\lambda $, as a function of (a) \gammaRatio, (b) \rgg\ and (c) \rvv\ for the combination with the on-shell signal strength measurement.
The dotted curves correspond to the one and two standard deviation confidence intervals on the measurement obtained using the Neyman construction.}
\end{figure}

%
%
%
%
%
%
%
%
%

% End of text imported from the .//sections/results.tex input file

\FloatBarrier
\section{Conclusion}
%

% The next lines are included from the .//sections/conclusion.tex input file
This Letter presents a search for off-shell Higgs boson production in the \zzfourl~and the \zztwoltwov~final states with the ATLAS detector using 139~\ifb~of $pp$ collision data. The search is optimised for sensitivity to both the ggF- and EW-induced signal, and a simultaneous fit to all SRs and CRs is performed to extract the signal contribution. No deviations from the SM prediction are observed. The data reject the background-only hypothesis with an observed (expected) significance of 3.3$\sigma$ (2.2$\sigma$).
The observed (expected) upper limit at 95\% confidence interval on the signal strength \muOFF~is found to be 2.4 (2.6). A combination with the on-shell Higgs boson measurement gives a measured total width of the Higgs boson~of $4.4^{+3.0}_{-2.2}$ \MeV, and the observed (expected) upper limit on the total width is found to be 10.1 (10.8)~\MeV~at 95\% CL.
These results are compatible with those previously obtained by the CMS experiment.
Together with that result, this means that both experiments have observed evidence of off-shell Higgs boson production.

% End of text imported from the .//sections/conclusion.tex input file

\FloatBarrier
\section*{Acknowledgements}
%

% The next lines are included from the .//acknowledgements/Acknowledgements.tex input file
%
%

%
%

We thank CERN for the very successful operation of the LHC, as well as the
support staff from our institutions without whom ATLAS could not be
operated efficiently.

We acknowledge the support of
ANPCyT, Argentina;
YerPhI, Armenia;
ARC, Australia;
BMWFW and FWF, Austria;
ANAS, Azerbaijan;
CNPq and FAPESP, Brazil;
NSERC, NRC and CFI, Canada;
CERN;
ANID, Chile;
CAS, MOST and NSFC, China;
Minciencias, Colombia;
MEYS CR, Czech Republic;
DNRF and DNSRC, Denmark;
IN2P3-CNRS and CEA-DRF/IRFU, France;
SRNSFG, Georgia;
BMBF, HGF and MPG, Germany;
GSRI, Greece;
RGC and Hong Kong SAR, China;
ISF and Benoziyo Center, Israel;
INFN, Italy;
MEXT and JSPS, Japan;
CNRST, Morocco;
NWO, Netherlands;
RCN, Norway;
MEiN, Poland;
FCT, Portugal;
MNE/IFA, Romania;
MESTD, Serbia;
MSSR, Slovakia;
ARRS and MIZ\v{S}, Slovenia;
DSI/NRF, South Africa;
MICINN, Spain;
SRC and Wallenberg Foundation, Sweden;
SERI, SNSF and Cantons of Bern and Geneva, Switzerland;
MOST, Taiwan;
TENMAK, T\"urkiye;
STFC, United Kingdom;
DOE and NSF, United States of America.
In addition, individual groups and members have received support from
BCKDF, CANARIE, Compute Canada and CRC, Canada;
PRIMUS 21/SCI/017 and UNCE SCI/013, Czech Republic;
COST, ERC, ERDF, Horizon 2020 and Marie Sk{\l}odowska-Curie Actions, European Union;
Investissements d'Avenir Labex, Investissements d'Avenir Idex and ANR, France;
DFG and AvH Foundation, Germany;
Herakleitos, Thales and Aristeia programmes co-financed by EU-ESF and the Greek NSRF, Greece;
BSF-NSF and MINERVA, Israel;
Norwegian Financial Mechanism 2014-2021, Norway;
NCN and NAWA, Poland;
La Caixa Banking Foundation, CERCA Programme Generalitat de Catalunya and PROMETEO and GenT Programmes Generalitat Valenciana, Spain;
G\"{o}ran Gustafssons Stiftelse, Sweden;
The Royal Society and Leverhulme Trust, United Kingdom.

The crucial computing support from all WLCG partners is acknowledged gratefully, in particular from CERN, the ATLAS Tier-1 facilities at TRIUMF (Canada), NDGF (Denmark, Norway, Sweden), CC-IN2P3 (France), KIT/GridKA (Germany), INFN-CNAF (Italy), NL-T1 (Netherlands), PIC (Spain), ASGC (Taiwan), RAL (UK) and BNL (USA), the Tier-2 facilities worldwide and large non-WLCG resource providers. Major contributors of computing resources are listed in Ref.~\cite{ATL-SOFT-PUB-2023-001}.

%
%

% End of text imported from the .//acknowledgements/Acknowledgements.tex input file

%

%
%
%
%
%
%
%
\clearpage
\printbibliography
\clearpage
% ATLAS Collaboration author list
% Reference date of HIGG-2018-32 is 2022-10-26
% Author list last updated on date 20-DEC-23
% Data extracted on 20-Dec-2023 for paper reference HIGG-2018-32
% at 10:50am
 
\begin{flushleft}
\hypersetup{urlcolor=black}
{\Large The ATLAS Collaboration}

\bigskip

\AtlasOrcid[0000-0002-6665-4934]{G.~Aad}$^\textrm{\scriptsize 102}$,
\AtlasOrcid[0000-0002-5888-2734]{B.~Abbott}$^\textrm{\scriptsize 120}$,
\AtlasOrcid[0000-0002-1002-1652]{K.~Abeling}$^\textrm{\scriptsize 55}$,
\AtlasOrcid[0000-0002-8496-9294]{S.H.~Abidi}$^\textrm{\scriptsize 29}$,
\AtlasOrcid[0000-0002-9987-2292]{A.~Aboulhorma}$^\textrm{\scriptsize 35e}$,
\AtlasOrcid[0000-0001-5329-6640]{H.~Abramowicz}$^\textrm{\scriptsize 151}$,
\AtlasOrcid[0000-0002-1599-2896]{H.~Abreu}$^\textrm{\scriptsize 150}$,
\AtlasOrcid[0000-0003-0403-3697]{Y.~Abulaiti}$^\textrm{\scriptsize 117}$,
\AtlasOrcid[0000-0003-0762-7204]{A.C.~Abusleme~Hoffman}$^\textrm{\scriptsize 137a}$,
\AtlasOrcid[0000-0002-8588-9157]{B.S.~Acharya}$^\textrm{\scriptsize 69a,69b,p}$
% ...
\bigskip
\\

% ...
$^{29}$Physics Department, Brookhaven National Laboratory, Upton NY; United States of America.\\
% ...
$^{35}$$^{(a)}$Facult\'e des Sciences Ain Chock, R\'eseau Universitaire de Physique des Hautes Energies - Universit\'e Hassan II, Casablanca;$^{(b)}$Facult\'{e} des Sciences, Universit\'{e} Ibn-Tofail, K\'{e}nitra;$^{(c)}$Facult\'e des Sciences Semlalia, Universit\'e Cadi Ayyad, LPHEA-Marrakech;$^{(d)}$LPMR, Facult\'e des Sciences, Universit\'e Mohamed Premier, Oujda;$^{(e)}$Facult\'e des sciences, Universit\'e Mohammed V, Rabat;$^{(f)}$Institute of Applied Physics, Mohammed VI Polytechnic University, Ben Guerir; Morocco.\\
% ...
$^{55}$II. Physikalisches Institut, Georg-August-Universit\"{a}t G\"ottingen, G\"ottingen; Germany.\\
% ...
$^{69}$$^{(a)}$INFN Gruppo Collegato di Udine, Sezione di Trieste, Udine;$^{(b)}$ICTP, Trieste;$^{(c)}$Dipartimento Politecnico di Ingegneria e Architettura, Universit\`a di Udine, Udine; Italy.\\
% ...
$^{102}$CPPM, Aix-Marseille Universit\'e, CNRS/IN2P3, Marseille; France.\\
% ...
$^{117}$Department of Physics, New York University, New York NY; United States of America.\\
% ...
$^{120}$Homer L. Dodge Department of Physics and Astronomy, University of Oklahoma, Norman OK; United States of America.\\
% ...
$^{137}$$^{(a)}$Departamento de F\'isica, Pontificia Universidad Cat\'olica de Chile, Santiago;$^{(b)}$Millennium Institute for Subatomic physics at high energy frontier (SAPHIR), Santiago;$^{(c)}$Instituto de Investigaci\'on Multidisciplinario en Ciencia y Tecnolog\'ia, y Departamento de F\'isica, Universidad de La Serena;$^{(d)}$Universidad Andres Bello, Department of Physics, Santiago;$^{(e)}$Instituto de Alta Investigaci\'on, Universidad de Tarapac\'a, Arica;$^{(f)}$Departamento de F\'isica, Universidad T\'ecnica Federico Santa Mar\'ia, Valpara\'iso; Chile.\\
% ...
$^{150}$Department of Physics, Technion, Israel Institute of Technology, Haifa; Israel.\\
$^{151}$Raymond and Beverly Sackler School of Physics and Astronomy, Tel Aviv University, Tel Aviv; Israel.\\
% ...
$^{a}$ Also Affiliated with an institute covered by a cooperation agreement with CERN.\\
$^{b}$ Also at An-Najah National University, Nablus; Palestine.\\
$^{c}$ Also at Borough of Manhattan Community College, City University of New York, New York NY; United States of America.\\
$^{d}$ Also at Bruno Kessler Foundation, Trento; Italy.\\
$^{e}$ Also at Center for High Energy Physics, Peking University; China.\\
$^{f}$ Also at Center for Interdisciplinary Research and Innovation (CIRI-AUTH), Thessaloniki; Greece.\\
% ...
$^{p}$ Also at Department of Physics, King's College London, London; United Kingdom.\\
% ...

\end{flushleft}

% Created with Glance <Atlas.Glance@cern.ch>

%

%
%
\clearpage

\end{document}